\documentclass[a4paper,12pt]{aa}
\usepackage{praeamble}
\usepackage{orcidlink}

\begin{document}

\title{Where did all the Little Red Dots go?}
\subtitle{The abundance of LRD analogues among objects with broad lines at $z<0.35$}
    


\author{L.~Seyberlich\orcidlink{0009-0009-5402-2727}\inst{1,2}\corrauth{lasse.seyberlich@uni-hamburg.de} \and S.~E.~I.~Bosman\orcidlink{0000-0001-8582-7012}\inst{1,3}\email{bosman@thphys.uni-heidelberg.de}}

\institute{Institut für Theoretische Physik, Universität Heidelberg, Philosophenweg 12, 69120, Heidelberg, Germany \and
\textcolor{black}{Hamburger Sternwarte, Universität Hamburg, Gojenbergsweg 112, 21029 Hamburg, Germany} \and
Max Planck Institut für Astronomie, Königstuhl 17, 69117, Heidelberg, Germany}

\date{Received date / Accepted date }

\abstract{One of the most puzzling discoveries since the launch of the \textit{James Webb Space Telescope} is a numerous population of sources at high redshift nicknamed ``Little Red Dots'' (LRDs). Characterized by broad Balmer emission lines, compact morphologies and v-shaped spectral energy distributions (SEDs) in the rest-frame optical, the nature of LRDs remains elusive. A major challenge is the high redshift nature of LRDs, which makes observations very expensive and difficult. Finding local LRD analogues is a crucial step in unravelling the physical processes giving rise to LRDs, as cheaper observations with many facilities would become available.
In this paper, we conduct a search for local LRD analogues by starting from an unbiased parent catalogue of all SDSS objects with broad H$\alpha$ or H$\beta$ emission lines (both galaxies and quasars) at $0 < z < 0.35$, complemented by archival photometry and spectroscopy. We find that among broad-line objects, $19\%$ show a v-shaped SED. 
We then select objects whose SED is well described by a modified black body (MBB), and identify nine local LRD analogues. \textcolor{black}{After accounting for selection limitations we get an abundance of $0.08\%$ of all objects with broad lines.} 
Among only objects which match the typical continuum and emission line luminosities of LRDs, $5\%$ display a v-shape and $0.37\%$ display a MBB-like SED.
We recover one previously-identified local LRD analogue (the so-called ``Egg''), validating our methodology. Among the other eight analogues, one (J0302$-$0101) shows all the hallmarks of an LRD despite being visibly embedded in the core of a galaxy; another object is ambiguous, and the remaining six may have alternative explanations for their peculiar SEDs. 
Our study highlights that LRDs are extremely rare at $z < 0.35$; we find very few objects with broad lines which have LRD-like SEDs, even without attempting to match the compactness of LRDs.
}



\maketitle
\nolinenumbers

\section{Introduction}\label{sec:intro}
The emergence of the \textit{James Webb Space Telescope} (\textit{JWST}) \citep{Gardner06} brought a puzzling population of objects to light.
``Little Red Dots'' (LRDs) were serendipitously discovered in several different studies while looking for other categories of high redshift objects, such as  active galactic nuclei (AGN) or massive galaxies \citep{Harikane23,Labbe23,Matthee24b,Barro24,Maiolino24}. Their unusual properties suggest that they harbour AGN powered by supermassive massive black holes (SMBHs) at $z>8$, which may provide vital clues to the open question of the origin of the first SMBHs \citep{Fan_review}. 

One obstacle to understanding LRDs that, at the time of writing, there is no consensus on a precise definition of these objects. 
Several studies have explicitly searched for LRDs and other red and compact sources in different fields and with different methods. 
They are generally identified through three physical properties with specific cut-off differing between authors. At high redshift, these properties co-occur, suggesting a potential common physical origin. 
These three properties are:
\begin{enumerate}
	\item a ``v-shaped'' spectral energy distribution (SED);
	\item a compact rest-frame optical morphology; and
	\item broad Balmer line emission.
\end{enumerate}


Multiple challenges arise when trying to understand the origin of these features of LRDs. First, the presence of broad Balmer emission lines is a telltale sign of an AGN. 
However, the rest-frame optical emission of AGNs is described by a blue power law arising from an accretion disc.
Explaining the red optical colours of LRDs (the v-shape) would then require a substantial amount of dust, which absorbs the UV light more than the optical, turning the observed spectrum red.
The light absorbed by the dust is then re-emitted in the NIR as thermal radiation.
This would result in a strong rise in the NIR of such an obscured AGN.
However, the SED of LRDs in the NIR flattens or even declines \citep{Williams24,PerezGonzalez24,Akins24,Leung25,Delvecchio25,Ronayne25}, contrary to bright early quasars which unambiguously host SMBHs \citep{Bosman24,Bosman25}. 
Furthermore, the v-shape of LRDs always has the same inflection point near the Balmer limit \citep{Setton24}, suggesting that this is an intrinsic feature.
If this v-shape was to be produced by the AGN (via scattered light and an obscured component), one would expect varying inflection points.
Additionally, the clustering of galaxies around LRDs sometimes hints at lower halo masses than expected for quasars at the same redshifts \citep{Arita24,LinX25c}, with however some exceptions \citep{Schindler24}.
In addition, if LRDs are bright AGN with dust obscuration, then detections of X-rays would be expected $-$ this is occasionally seen \citep{Kocevski25,Hviding26,Fu25} but the bulk of LRDs are X-ray-weak \citep{Ananna24,Yue24,SacchiBogdan25,Tortosa26}. 
Similarly, in the radio wavelengths, studies only find upper limits that are only compatible with radio-weak (or radio-quiet) AGNs \citep{Akins24,Perger25b,Gloudemans25}. 
These considerations have led to scepticism for an interpretation of LRDs as purely obscured AGNs.

Interpretations of LRDs as purely galactic, without the presence of AGNs, are also untenable. While the SED v-shape can be produced by an old stellar population, the broad lines and the compact appearance of LRDs are hard to explain, as well as missing cold dust in the FIR \citep{Labbe25,Casey25,Xiao25}.
Interestingly, the compact morphology of LRDs is more prominent in the rest-optical, while the rest-UV shows more extended asymmetric emission for some objects \citep{Kokorev24,Rinaldi25b,ZhangY25}. Nevertheless, the stellar masses inferred from treating LRD SEDs as purely galactic would be in tension with expectations from cosmology \citep{Labbe23}.

Some newer explanations for LRDs invoke unusual accretion processes to explain their observed properties.
To reproduce the observed SED inflection point around the Balmer limit without the need for a stellar contribution, models invoke a dense gas surrounding an AGN in configurations reminiscent of a  ``Black Hole Star'' \citep{InayoshiMaiolino25,Ji25a,deGraaff25,Naidu25}, that can lead to extreme Balmer breaks as well as giving rise to narrow Balmer absorption, which is observed in $\gtrsim 30\%$ of LRDs \citep{Matthee24b, WangB25b, Naidu25,deGraaff25,Rusakov25,Matthee26}.
Additionally, gas instead of dust attenuation does not produce strong emission in the NIR and can explain the X-Ray weakness.
This hypothesis has given rise to a useful empirical fitting form for the SEDs of LRDs: 
\citet{deGraaff25b} perform modified black body (MBB) fitting to describe the red optical and flattening in the NIR on their constructed sample.
They find that LRDs are well described by  MBBs with typical temperatures around $T=5000\,\mathrm{K}$, which can be interpreted as the ``surface temperature'' of a Black Hole Star-like structure. 
Similar results are obtained by \citet{Barro26} and \citet{Asada26}.

The implemented geometry of the dense gas varies between different models, as does the interpretation of the attenuated ``central engine'' \citep[e.g.][]{BegelmanDexter25,Chisholm26,Zwick25,Kido25,LiuH25,LinX25b,Pang26,Umeda25,Inayoshi25b, ZhangC26}.
Some of these models imply super-Eddington accretion, the feasibility of which is being actively investigated (e.g.~\citealt{LiuH25,Secunda25}).
Disentangling these models is non-trivial and will require more consensus from observations. 


An important obstacle to figuring out the physics of LRDs comes from the lack of a precise definition and varying selection methods.
There are three primary selection strategies, based on either:
\begin{enumerate}
	\item photometric colours \citep{Labbe25,Kokorev24, Akins24, Barro24b, Barro26};
	\item photometric continuum slopes \citep{Kocevski25,Zhuang24b}; or
	\item spectroscopic continuum slopes \citep{Hviding25, deGraaff25b}.
\end{enumerate}
While photometric searches are very accurate in identifying LRDs, they are highly incomplete \citep{Hviding25}, making number density estimates very uncertain.
Some studies suggest that LRD selections are only picking up a fraction of a broader population of LRDs or even just AGN \citep{Rinaldi26,Billand26}.
These differing selections could potentially dilute LRD samples and lead to varying observed properties between studies.
For example, while most studies find no significant AGN-like variability in LRDs \citep{KokuboHarikane25,ZhangZ25a,Tee25,LiuZ26}, strong variability may have been observed in another LRD \citep{Lambrides26} and in some observations utilizing lensing which show long-term variability \citep{ZhangZ25b,Furtak25,Ji25a}.

LRDs are found predominantly at high redshifts ($z\sim 4-9$) and their number density drops steeply below $z \approx 4$ \citep{Kocevski25,Ma25, Euclid25, Tanaka25}.
However, searches for finding local analogues have been carried out, as local LRDs could be an important laboratory for unravelling the nature of LRDs.
To this date, only a few rare local analogues have been found that satisfy LRD selection criteria.

\citet{LinX25b} transpose high-$z$ LRD spectra to $z = 0.0$-$0.5$ and calculate synthetic broad band photometry.
This is compared against Sloan Digital Sky Survey \citep[SDSS][]{SDSS00} and \textit{Galaxy Evolution Explorer} \citep[\textit{GALEX};][]{GALEX05} photometry.
They then require a compact morphology and several properties of oxygen emission lines to remove contaminants which show a Balmer break such as stars, quiescent galaxies and post-starburst galaxies.
To ensure a red optical SED, they demand that the continuum near H$\alpha$ be stronger than around [O~{\small{III}}]$\lambda 5007$, and at the same time that the continuum around [O~{\small{III}}]$\lambda 5007$ be stronger than around [O~{\small{II}}]$\lambda 3730$.
Of their remaining 40 sources they present three as ``local LRDs'' and disregard the others as they show high [N~{\small{II}}] or Mg~{\small{II}} emission that they deem more likely to be produced by type-I AGN.
One of their local LRDs, nicknamed ``The Egg'', shows a blueshifted Balmer absorption feature.
This object was found independently by \citet{Ji25b} (nicknaming  it ``Lord of LRDs''), who also confirm its extreme X-Ray weakness.
``The Egg'' shows strong Na D, K ~{\small{I}}, and Ca ~{\small{II}} triplet absorption lines and a series of [Fe~{\small{II}}] emission lines.
\citet{Rodriguez25} investigate the radio emission of ``The Egg'', but cannot find an associated source.
In contrast, one of the other two sources from \citet{LinX25b} shows radio signatures typical for optically thin synchrotron emission.
Furthermore, the three sources are showing no significant signs of variability on a baseline of $>10\,\mathrm{yrs}$ \citep{Burke25}.
These three objects are generally accepted by the literature as local LRDs/local LRD analogues.

There are some additional claims for local LRDs: 
\citet{LinR25,LinR26} claim green pea galaxies to be local LRDs, while \citet{ChenX25} claim an unresolved core in an extended starburst and ultra-luminous IR galaxy as an LRD-like galaxy.
\citet{Ding26} discovered five sources from the Dark Energy Spectroscopic Instrument \citep[DESI][]{DESI16} DR1 data that meet the high redshift LRD selections.
They measure continuum shapes from the spectra to determine a v-shape, demand compact morphology and a similar oxygen cut as \citet{LinX25b} to exclude other objects with prominent Balmer breaks.
However, they point out that, while their sources fulfil the LRD selection riteria (compactness, broad Balmer lines and v-shaped SED), they occupy a different space in the Baldwin, Phillips \& Terlevich (BPT) diagram \citep{BPT81} than LRDs and ``The Egg''.
Therefore, they caution that their sources could be physically related to high redshift LRDs, but a pure observational similarity is possible as well.

\citet{Park26} also perform a search in DESI DR1.
There is no overlap between the eight sources they find and the sources from \citet{Ding26}.
This is due to a different selection method: 
\citet{Park26} select their sources mainly on emission line properties, including a cut that demands weak [N~{\small{II}}] and [S~{\small{II}}] doublets.
The objects from \citet{Ding26} do not satisfy that demand.
\citet{LinX26} conducted a search in DESI DR1 and find 27 local LRD analogues with a strategy based on \citet{LinX25b}.
\citet{Shangguan26} find seven objects with Balmer line absorption through visual inspection of a representative and homogenous catalogue of broad line AGN \citep{Liu19}.
However, only some of their sources show signatures of a v-shape in their spectra.
\citet{Casey26} build a sample of $\sim 1325$ photometric local LRD candidates from the SDSS DR16 quasar catalogue \citep{Lyke20}.
In their v-shape selection they loosen the UV slope cut compared to high redshift searches, as stellar contribution is more prominent in the UV of LRDs and might change at lower redshift.
They do not make any cuts regarding compactness for their sample.


In conclusion, so far only a few local LRD analogues have been discovered. 
These analogues are mostly selected on the exact same definitions as high redshift LRDs.
While this certainly is a reasonable approach for finding local analogues, the selection criteria themselves could in principle have a redshift dependence.
Accounting for such changes and modifying the criteria accordingly as well as comparing the empirical parameter space of LRDs to local sources has not been done so far.
In this study, we approach the search for local LRDs from a homogenous parent sample objects at $z<0.35$ displaying broad emission lines, and we relax the requirement that the morphology of the source need be compact. In this way, we hope to identify objects of interest for future detailed studies, and to quantify the occurrence rate of LRD-like objects in the local Universe without the bias of morphology.

The rest of the paper is structured as follows. High redshift LRDs and local archival data (Sec. \ref{sec:data}) are used to build selection criteria accounting for redshift changes as well as the observational parameter space LRDs occupy (Sec. \ref{sec:sample_construction}), followed by a brief analysis of that sample in Sec. \ref{sec:sample_analysis}.
Lastly, we discuss our findings in Sec. \ref{sec:discussion} and conclude in Sec. \ref{sec:summary}.

Throughout this work we adopt a flat $\Lambda$CDM cosmology with $\Omega_m = 0.315$ and $h = 0.674$ \citep{PlanckColab20}.

\section{Data}\label{sec:data}
In this study, we aim to compare a sample of high-redshift LRDs to a general local AGN sample.
We obtained data for high-redshift LRDs from archival \textit{JWST} observations, and used a sample of local broad-line objects for the low-$z$ reference sample, which we complemented with archival observations from various surveys as necessary.

\subsection{Local Parent Sample}
The low-$z$ parent sample comes from \citet[][\citetalias{Liu19} hereafter]{Liu19}, who constructed a catalogue comprising of all objects in the Sloan Digital Sky Survey Seventh Data Release \citep[SDSS DR7;][]{SDSSDR7} which possess broad lines. 

\citetalias{Liu19} start from all spectra in SDSS DR7 at $z<0.35$ which are classified either as ``galaxies'' or ``quasars'', meaning $854,664$ ``galaxies'' and $11,638$ ``quasars''. The maximum redshift is set to ensure that the H$\alpha$ line is well covered by the wavelength range of the SDSS spectrograph ($3800$-$9200\,\mathrm{\AA}$).
To remove stellar absorption lines and continuum from potential hosts, they perform an ensemble learning independent component analysis technique \citep[EL-ICA;][]{Lu06} to decompose stellar and nuclear components.
Next, in order to determine whether broad Balmer lines are present,  
\citetalias{Liu19} fit the Balmer lines as well as the close-by [N~{\small{II}}]$\lambda \lambda 6548,6583$ and [S~{\small{II}}]$\lambda \lambda 6716,6731$ doublets in the H$\alpha$ region and the [O~{\small{III}}]$\lambda \lambda 4960,5007$ doublet close to H$\beta$.
Instead of defining a fixed broad line threshold, the criterion 
an object is classified as presenting broad emission lines if the inclusion of a broad component significantly improves the fit. 

For all objects classified as containing broad H$\alpha$ and/or broad H$\beta$ lines, \citetalias{Liu19} report the line fluxes and FWHMs, as well as derived properties including bolometric luminosities $L_{\mathrm{bol}}$ and black hole masses $M_{\mathrm{BH}}$. 
This final catalogue (hereafter called the \citetalias{Liu19} catalogue) consists of $14\,584$ AGN and galaxies with broad lines, regardless of any other observed properties. 
For the purposes of finding low-reshift LRD analogues, the \citetalias{Liu19} catalogue is preferable as a parent sample compared to newer quasar catalogues (e.g.~\citealt{Lyke20}) because of \citetalias{Liu19}'s inclusion of sources classified as galaxies.
This is relevant for an LRD search as LRDs show properties typical for galaxies - like a Balmer break. 
Furthermore, this means that the \citetalias{Liu19} catalogue includes many low luminosity AGNs, that are missed by the SDSS classification and are not part of SDSS quasar catalogues.

Additionally, \citetalias{Liu19} provide photometric data for the sources in the catalogue.
SDSS point spread function (PSF) and Petrosian photometry in \textit{ugriz}-bands \citep{SDSSDR7} are given and throughout this work PSF photometry is used.
In the infrared, observations from the \textit{Wide-field Infrared Survey Explorer} \citep[\textit{WISE};][]{WISE10} with the filters W1, W2, W3 and W4 and the Two Micron All Sky Survey \citep[2MASS;][]{2MASS06} with \textit{J}, \textit{H} and $K_S$ are used.
In the UV, GALEX FUV and NUV bands are provided in the \citetalias{Liu19} catalogue.

\subsection{High Redshift LRD Sample}
The sample of high redshift LRDs is taken from \citet[][dG25 hereafter]{deGraaff25b}.
There are \textcolor{black}{181} sources in \citetalias{deGraaff25b} of which \textcolor{black}{146} are unique sources.
In the following only these 116 sources are considered.
For a sub-set of \textcolor{black}{46} unique sources, \citetalias{deGraaff25b} perform fitting involving a modified black-body function; whenever we are performing similar black-body fits in the low-$z$ sample, we only compare to those \textcolor{black}{46} sources.

\subsection{Auxiliary Data}
In addition to the information provided by the \citetalias{Liu19} catalogue, some complementary observational data is gathered for the low-$z$ objects.

The analysis in \citetalias{Liu19} is based on SDSS spectra, but their data release contains only the derived observational properties and not the spectra themselves. 
Therefore, we retrieve the spectra for all sources used in \citetalias{Liu19} from the SDSS Science Archive Server\footnote{\url{https://data.sdss.org/sas}} (SAS).
Additionally, we check whether any objects have since been re-observed within SDSS by the extended Baryon Oscillation Spectroscopic Survey \citep[eBOSS;][]{EBOSS16}. 
If this is the case, these spectra are obtained as well, resulting in 1101 additional eBOSS spectra.

For some individual sources, observations from the Spectro-Photometer for the History of the Universe, Epoch of Reionization, and Ices Explorer \citep[SPHEREx;][]{SPHEREx14} are used.
In principle, SPHEREx observations cover the entire sky, but due to the recency of its public data release and ongoing challenges in retrieving data for large catalogues of objects, we are only able to use SPHEREx data for individual objects of particular interest.
For these objects the data is obtained from the NASA/IPAC Infrared Science Archive 
\footnote{\url{https://irsa.ipac.caltech.edu}} (IRSA).

\section{Sample Selection} \label{sec:sample_construction}
Selection criteria are constructed to resemble the population of LRDs as closely as possible, while accounting for possible changes due to different redshift.
This study only uses the empirical observed properties of LRDs 
regardless of the physical interpretation of these properties.

\subsection{Standard Selection Criteria}
First, the criteria used for LRD selection at high redshift are considered.
These are: a v-shaped SED, compactness, and broad Balmer emission lines.

\subsubsection{V-Shape}\label{sec:v-shape}
In order to select low-$z$ analogues based on the v-shape of their SEDs, we first determine power-law slopes for the rest-frame UV and rest-frame optical wavelengths. 
In the optical, a power law $f_\lambda = A_{\mathrm{OPT}}\cdot \lambda^{\beta_{\mathrm{OPT}}}$ is fitted to the spectra of the \citetalias{Liu19} catalogue between $3645 - 7000\,\mathrm{\AA}$, while avoiding the regions of emission lines H$\gamma$, H$\beta$ and H$\alpha$. 
For the rest-frame UV slope, $f_\lambda = A_{\mathrm{UV}}\cdot \lambda^{\beta_{\mathrm{UV}}}$, an analogues approach is taken for wavelengths shorter than the Balmer limit.
This is complicated due to the fact that at $z < 0.04$ the Balmer break falls out of the detector range of the SDSS spectrograph; even for sources in the higher redshift range of the parent sample, only very few data points exist blueward of the Balmer break.
Additionally, the blue end of the spectra has higher spectral uncertainties, sometimes resulting in poor constraints on the UV power law.

Therefore, an improved approach is taken. 
We complement the spectra with photometric data including the \textit{GALEX} NUV and FUV  and SDSS \textit{u} bands.
Additionally, if the 20 bluest data-points of the spectrum are blueward of $3645\,\mathrm{\AA}$, they are binned and added as an additional point for the power-law fit.
The spectral binning is done to ensure that spectral data does not dominate the fitting procedure compared to photometry.

If only two points are available, the direct analytical expression: 
\begin{equation}
	\beta_{\mathrm{UV}} = \frac{0.4 (m_1 - m_2)}{\log (\lambda_2/\lambda_1)} - 2\,\mathrm{,}
\end{equation}
is used for the calculation instead of a fitting routine.

In the event that there is only one datapoint after this selection (i.e.~no complementary photometry exists), the 20 bluest spectral points are used for the fitting.
This is necessary for only 62 sources.

\citet{Hviding25} demonstrate that using photometry for the slope determination can result in biased colours as line emission can contaminate the fitting.
However, this is mostly related to the red optical part of the v-shape SEDs of LRDs and should be a lesser issue in the blue/UV side.

\begin{figure}
	\resizebox{\hsize}{!}{\includegraphics{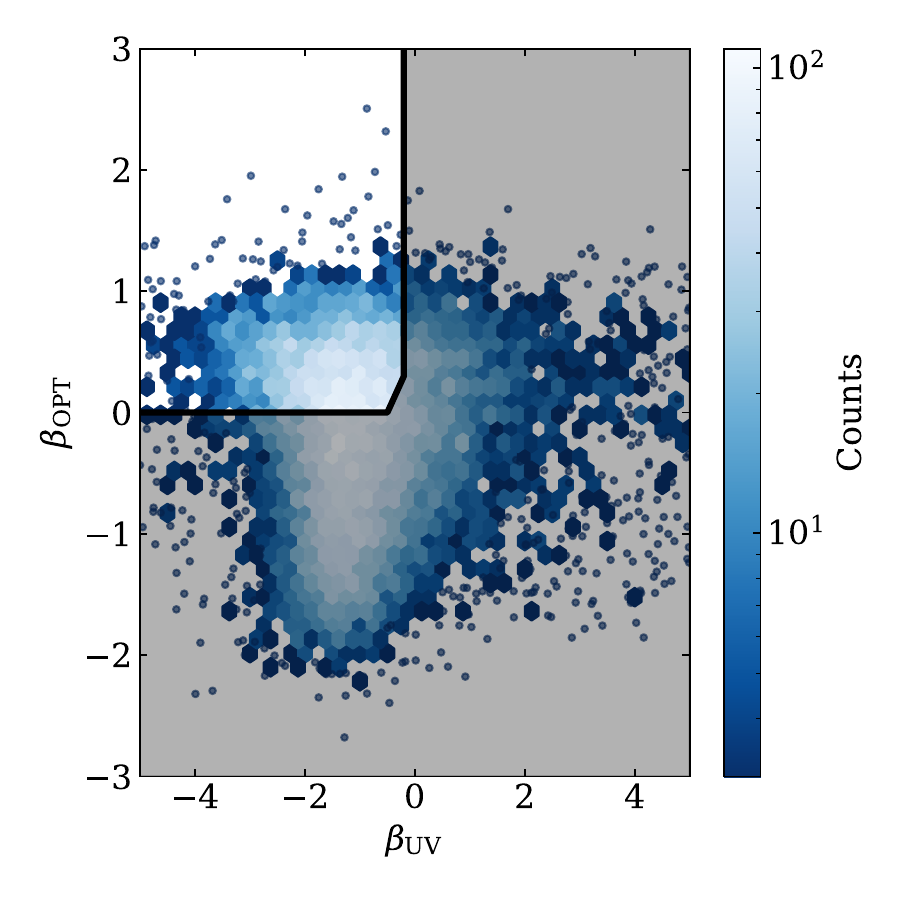}}
	\caption{The measured slopes in the UV ($\beta_{\mathrm{UV}}$) and in the optical ($\beta_{\mathrm{OPT}}$). The black lines indicate the colour cuts implemented. The left upper corner represents the objects considered after the selection, which satisfy the criterion of ``v-shaped'' SED.}
	\label{fig:slope_uv_opt}
\end{figure}

For the classification of a v-shape, the criteria of \citet{Hviding25} are applied.
This means $\beta_{\mathrm{UV}} < -0.2$, $\beta_{\mathrm{OPT}} > 0$ at $2\sigma$ detection and $\beta_{\mathrm{OPT}} - \beta_{\mathrm{UV}} > 0.5$. 
Additionally, the amplitude of the fits are forced to be positive.
The resulting distribution of optical and UV slopes is shown in Fig.~\ref{fig:slope_uv_opt}.
This v-shape selection is fulfilled by 2778 sources.

\subsubsection{Compactness}
Compactness is a core criterium in most studies which have searched for LRDs so far.
With their spectroscopic sample, \citet{Hviding25} demonstrate that selecting sources with a point source morphology paired with a v-shaped SED yields broad lines ubiquitously.
However, this is a statement about high redshift sources alone.
Due to the limited knowledge of the physical properties of LRDs, it is not clear \textit{a priori} that this should hold for low redshift LRD analogues as well.
In particular, the potential host galaxies of LRDs may be different cosmic time due to galaxy evolution, or may be more frequently missed at early times to surface brightness dimming, which scales as $(1+z)^{-4}$ \citep{Calvi14}.
Following this line of thought, \citet{Rinaldi25} identified an extended IR galaxy at $z = 2.0145$ which showed a v-shape SED in its nucleus and simulated its photometry if it were moved to $z = 7$.
The resulting object is a compact red dot, demonstrating the importance of the surface brightness dimming effect.

In conclusion, we opt to drop the compactness criterion in our search for local LRD analogues, as it is possibly a high-redshift effect rather than an intrinsic property of LRDs. 

\subsubsection{Broad Lines}
The third major selection criterion for LRDs are broad Balmer lines.
The parent sample from the \citetalias{Liu19} catalogue is a sample of broad line AGN and galaxies.
Therefore, broad Balmer lines are present by definition in all sources considered.
However, the specific selection criteria of the \citetalias{Liu19} catalogue differ slightly from the more common cut of $\mathrm{F\!W\!H\!M} > 1000\,\mathrm{km\,s^{-1}}$, which is used in LRD selections.
\citetalias{Liu19} define an object as having broad lines if fitting its spectrum with a broad component significantly improves the fit compared to a fit without a broad component.
This means that some sources in the catalogue have effective total $\mathrm{F\!W\!H\!M} < 1000\,\mathrm{km\,s^{-1}}$.
However, this is only a very small subset of 219 objects. We therefore impose no direct cuts on the FWHM.

\subsection{Additional Selection Criteria}
The selection of applicable standard criteria for LRDs yields a reduced sample of 2778 objects, resulting in a fraction of $\sim19\%$ of the broad-line parent sample.
However, as compactness is not demanded, the similarity of this sample to LRDs is highly uncertain.
Therefore, additional observational parameters of LRDs are considered to find analogues objects within this reduced sample.

For all 2778 objects, SDSS or eBOSS spectra are available.
We use these spectra to derive other quantities for which LRDs show distinct behaviours, identified by 
\citetalias{deGraaff25b}. 
We chose this study since their LRD sample is constructed with the goal of purity over completeness in order to obtain a better understanding of the physical properties of LRDs without worrying about contaminants. 
This is especially important as one might expect more contaminants at lower redshift, both due to the lower number density of LRDs,  and also due to the existence of physically reasonable scenarios that could mimic an LRD (e.g.~a Balmer break from an old (Gyr) stellar population combined with a blue power law from the accretion disk of an AGN). 
We therefore derive several quantities from \citetalias{deGraaff25b}'s analysis for all 14584 objects in the \citetalias{Liu19} catalogue. 

\subsubsection{Continuum, Line and MBB Luminosities}
\citetalias{deGraaff25b} find a strong correlation between the luminosity of H$\alpha$ $L_{\mathrm{H}\alpha}$ and the continuum luminosity $L_{5100}$ at $5100\,\mathrm{\AA}$.

The \citetalias{Liu19} catalogue provides both of these quantities.
For $L_{\mathrm{H}\alpha}$ the narrow and broad line values are combined to stay consistent with \citetalias{deGraaff25b}.
$L_{5100}$ in \citetalias{Liu19} is derived from AGN-host decomposition.
However, using these values for $L_{5100}$ would bias the selection since for LRDs no such decomposition is performed when measuring $L_{5100}$.
Therefore, $L_{5100}$ is determined directly from the spectra.
We measure the continuum luminosity by averaging the flux around $5100\,\mathrm{\AA}$ in a $30\,\mathrm{\AA}$ window.
Errors are estimated by taking 1000 random Gaussian draws from the error spectrum.

\begin{figure}
	\resizebox{\hsize}{!}{\includegraphics{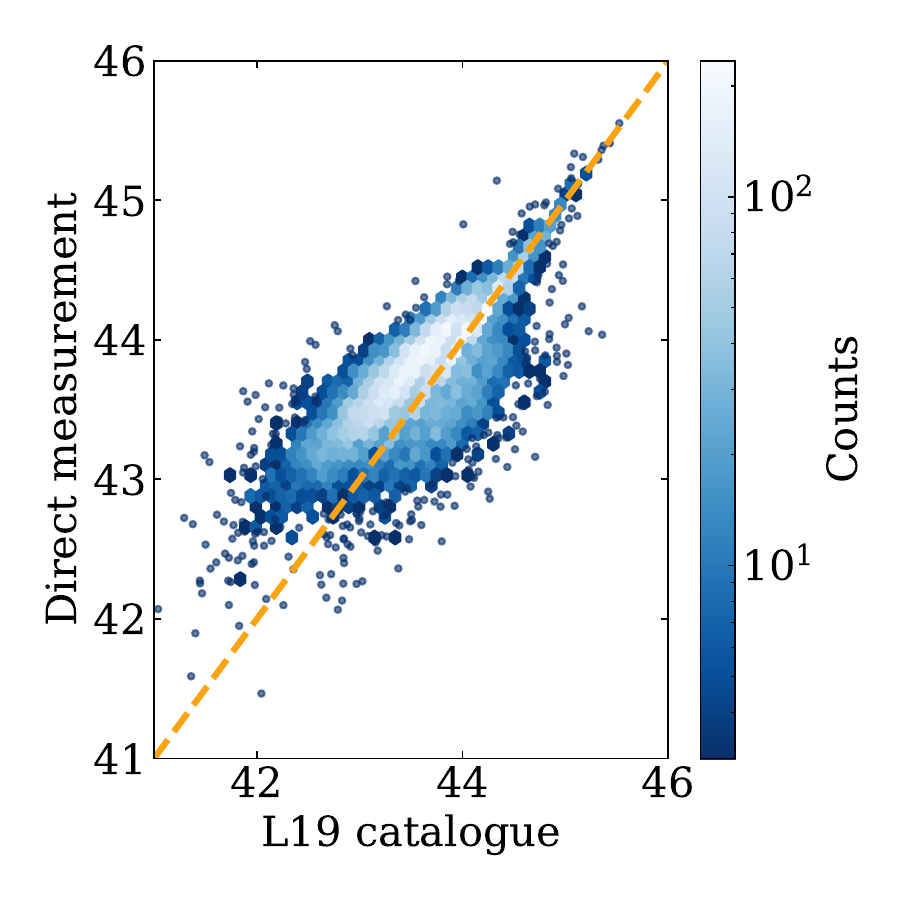}}
	\caption{$\log{L_{5100}}$ values reported in \citetalias{Liu19} compared to values measured directly from the spectra for all objects in the \citetalias{Liu19} catalogue. While \citetalias{Liu19} takes into account that some amount of the emission can come from the host, which is not included in their reported $L_{5100}$ value, the direct measurement takes all emission into account for $L_{5100}$ calculation. This is done to ensure consistency with the values measured for LRDs, where host decomposition is not performed, as contributions from a host are not clear a priori. The dashed yellow line indicates unity.}
	\label{fig:L5100_cal_vs_liu}
\end{figure}

Figure \ref{fig:L5100_cal_vs_liu} shows our measurements of $L_{5100}$ compared to the values from \citetalias{Liu19}. 
There is substantial scatter around the 1-to-1 line and a clear bias towards our values being larger, indicating the necessity of this process.
At the high-luminosity end the measurements are in better agreement, as expected since the AGN starts to dominate the spectrum compared to the host galaxy. 
For lower luminosity objects, some of the scatter may originate from the fact that \citetalias{Liu19} use the scaling relation from \citet{GreeneHo05} to estimate the continuum luminosity from H$\beta$ or H$\alpha$ for some of the objects in their catalogue. Some scatter may therefore be driven by scatter in the \citet{GreeneHo05} relation.

Figure \ref{fig:population_map} shows the sample from \citetalias{deGraaff25b} overlaid with the local sample.
As found by \citetalias{deGraaff25b}, LRDs follow a relation of $\log_{10} L_{5100} = \log_{10} L_{\mathrm{H}\alpha} + 0.8$.
Most objects in the local sample lie below this line, with only a few sources above.
Therefore, using a lower limit as a selection cut is sufficient $-$ we place this cut at $\log_{10} L_{5100} - \log_{10} L_{\mathrm{H}\alpha} < 1.4$.
This gives us 2384 objects from \citetalias{Liu19} that show LRD-like continuum and H$\alpha$ line luminosities.
Among these 2384 objects, 112 additionally show a v-shape, giving a fraction of $\sim5\%$.

\subsubsection{Modified Black Body Fitting}
One key finding from \citetalias{deGraaff25b} is the optical SEDs of LRDs can be described well with a modified black body function of the form 
\begin{equation}\label{eq:mbb_fnu}
	f_{\mathrm{MBB}} = A_{\mathrm{tot}} B_\nu(T) \left(\frac{\nu}{\nu_0}\right)^{\beta_{\mathrm{MBB}}}\,\mathrm{,}
\end{equation}
where $A_{\mathrm{tot}}$ is an amplitude, $B_\nu(T)$ the Planck function, $\nu_0 = c/5500\,\mathrm{\AA}$ a pivot frequency and $\beta_{\mathrm{MBB}}$ the exponent indicating the modification of the black body.
The parameter $\nu_0$ is introduced to reduce the dynamical range of $A_{\mathrm{tot}}$, but has no impact on the resulting values for $\beta_{\mathrm{MBB}}$ and $T$.
In addition, the parameter $\beta_{\mathrm{MBB}}$ effectively works as a broadening or narrowing of the black body.
$\beta_{\mathrm{MBB}} < 0$ broadens the black body, while $\beta_{\mathrm{MBB}} > 0$ narrows it.
If $\beta_{\mathrm{MBB}} = 0$ a unmodified black body is retrieved.

\begin{figure*}
    \includegraphics[width=17cm]{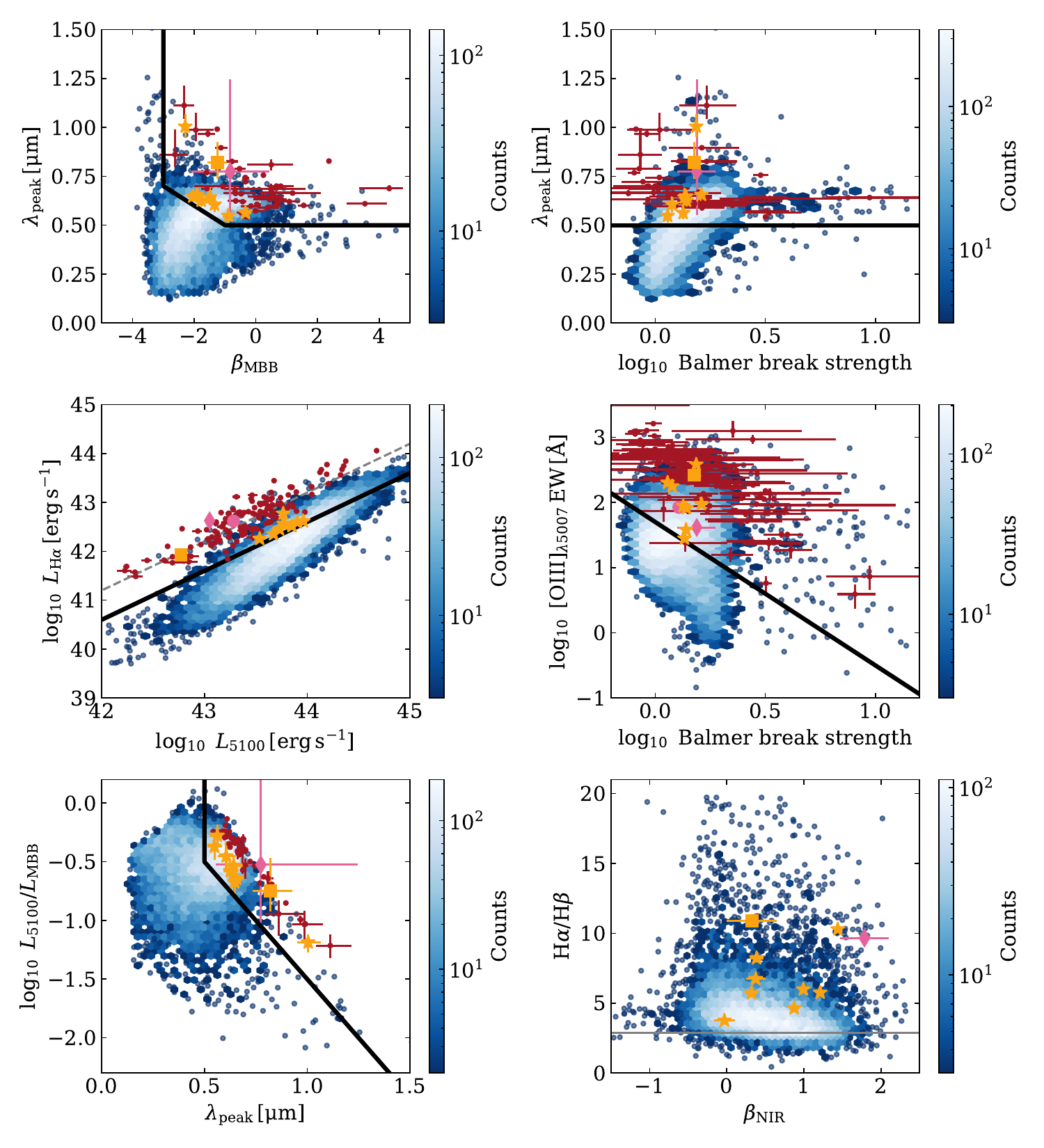}
	\caption{Diagnostic map of LRDs (little red dots), the local parent sample (blue points and distributions) and the final selection local LRD analogues (yellow stars). 
		\textit{Upper left:} $\beta_{\mathrm{MBB}}$ vs. $\lambda_{\mathrm{peak}}$. 
		\textit{Upper right:} $\log_{10} \text{BBS}$ vs. $\lambda_{\mathrm{peak}}$.
		\textit{Middle left:} $\log_{10} L_{5100}$ vs. $\log_{10} L_{\mathrm{H}\alpha}$.
		\textit{Middle right:} $\log_{10} \mathrm{EW}_{\lambda 5007}$ vs. $\log_{10} \text{BBS}$.
		\textit{Lower left:} $\lambda_{\mathrm{peak}}$ vs. $\log_{10} L_{5100}/L_{\mathrm{MBB}}$.
		\textit{Lower right:} $\beta_{\mathrm{NIR}}$ vs. $\mathrm{H}\alpha/\mathrm{H}\beta$. 
        \textcolor{black}{``The Egg'' is displayed with a yellow square, while the two missed objects discussed in Sec. \ref{sec:missed_objects} are shown in pink (J1047+0739 as a circle and J1022+0841 as a diamond) in all panels where that is possible.}
        The \textcolor{black}{dashed grey line in the middle left panel indicates the slope fitted to the high redshift sample in \citet{deGraaff25b}, the} solid grey line \textcolor{black}{in the lower right panel} indicates the theoretical expectations for case B recombination,  $\mathrm{H}\alpha/\mathrm{H}\beta = 2.86$ \textcolor{black}{and the black lines mark the selection cuts specified in Table \ref{tab:sample_criteria}.}}
	\label{fig:population_map}
\end{figure*}

Apart from best-fit values for $\beta_{\mathrm{MBB}}$ and $T$, one can also retrieve the peak wavelength $\lambda_{\mathrm{peak}}$ of the modified black body in $f_\lambda$, which \citetalias{deGraaff25b} found to be the better choice for analysis compared to $T$.
This is due to a known degeneracy between $\beta_{\mathrm{MBB}}$ and $T$, especially at low SNR \citep{Shetty09}.
Therefore, $\lambda_{\mathrm{peak}}$ is determined for the local parent sample as well. The top-left panel of 
Figure \ref{fig:population_map} shows that the local sample is mostly offset from the LRD sample, as the LRDs occupy a region which is at one end of the local distribution.
We empirically quantify this region with the selection cuts $\beta_{\mathrm{MBB}} > -3$, $\lambda_{\mathrm{peak}} > 0.5$ and $\lambda_{\mathrm{peak}} + 0.1 \cdot \beta_{\mathrm{MBB}} > 0.4$, as shown in Figure \ref{fig:population_map}.

Additionally, the ratio of the continuum luminosity to the integrated MBB luminosity ($L_\text{MBB}$) shows a strong anti-correlation with the peak wavelength in \citetalias{deGraaff25b}. 
We calculate $L_\text{MBB}$ in the low-$z$ sample by integrating the best-fit MBB fits and assuming the SDSS redshifts for the objects. We find that the anti-correlation between $L_{5100}/L_\text{MBB}$ and $\lambda_\text{peak}$ is a good discriminant of LRDs compared to the low-$z$ comparison sample (see bottom-left panel of Figure \ref{fig:population_map}). 
Motivated by this, a selection cut of $\log_{10} L_{5100}/L_{\mathrm{MBB}} + 2 \cdot \lambda_{\mathrm{peak}}  > 0.5$ is implemented together.

\subsubsection{Balmer-like Break}
A Balmer-like break is another prominent feature in LRDs.
The quantification of the Balmer break strength (BBS) used in this study is adopted from \citet{deGraaff25}.
The Balmer break is calculated by integrating the spectrum in two regions at $A =[3620,3720]\,\mathrm{\AA}$ and $B = [4000,4100]\,\mathrm{\AA}$ in $f_\nu$. 
The ratio $B/A$ is then computed.
Errors are estimated by taking 1000 random Gaussian draws from the spectral uncertainties.

Comparing the Balmer break strength to the peak wavelength (top right panel of Figure \ref{fig:population_map}) shows once more a significant overlap between high-$z$ LRDs and some local sources.
However, due to the large scatter in Balmer break strength in the high-$z$ LRD sample, a cut solely on this quantity is not informative in regard to the \citetalias{Liu19} catalogue.

\subsubsection{[O~{\small{III}}] Emission}
A further relation found by \citetalias{deGraaff25b} is the anti-correlation between the equivalent width of [O~{\small{III}}]$_{\lambda5007}$ and the Balmer break strength.
The middle right panel of Figure \ref{fig:population_map} shows a comparison of these quantities in the two samples.
As the LRDs overlap with the local sources for high values of at least one of the two quantities a diagonal lower limit is implemented as $\log_{10} \mathrm{EW}_{\lambda 5007} + 2.2 \cdot \log_{10} (\text{BBS}) > 1.7$.

\subsubsection{Other Parameters}
Apart from the aforementioned parameters, other properties are also typically found in LRDs. 
However, as they do not show clear discriminating power compared to the previously-discussed selection parameters, they are not used to define cuts.

First, LRDs often display large values of the Balmer decrement $\mathrm{H}\alpha/\mathrm{H}\beta$, potentially indicating obscuration of the broad line emitting region. LRDs display very wide scatter in the Balmer decrement \citep{deGraaff25b}.

In addition, LRDs have been reported to have a deficit of hot dust seen through a flat NIR SED slope (in $f_\nu$) \citep{Akins24, Williams24, Setton25, WangB25b}.
We evaluate the NIR slope by fitting a power law $f_\nu = A \cdot \nu^{\beta_{\mathrm{NIR}}}$ to the \textit{J}, \textit{H} and \textit{$K_s$} photometric bands of 2MASS and the W1 and W2 bands from \textit{WISE}.
All objects are not covered by all of these photometric bands; the fit is performed as long as at least two bands are available. 
For 26 object, less then two bands are available and the analysis cannot be conducted.
We find that low-$z$ objects occupy a broad range $-1\lesssim\beta_\text{NIR}\lesssim2$. However, similar information on individual LRDs is not available. So far, studies of LRDs at high redshift have only been able to analyse photometric stacks of sources.

The distribution of the Balmer decrement and the NIR slopes of the \citetalias{Liu19} catalogue is shown in the bottom right panel of Figure \ref{fig:population_map}.

\subsection{Final Selection}
Table \ref{tab:sample_criteria} lists all the selection criteria as well as the number of objects which are selected by each cut individually and the remaining objects after each consecutive cut.
While the v-shape selection cuts are done at the $2\sigma$ confidence level as in \citet{Hviding25}, the other cuts are implemented at the $1\sigma$ confidence level. 
After implementing all selection cuts, we find 9 objects in the parent sample that qualify as ``LRD-like'' in all categories, resulting in a fraction of $0.06\%$ of the homogenous broad-line parent sample at $z<0.35$.

\begin{table*}
	\caption{Criteria for the LRD candidate sample with the amount of objects selected per cut, the remaining objects after each cut \textcolor{black}{and the required significance of each cut. The additional selection cuts for the estimation of the abundance in Sec. \ref{sec:strictness} are given at the bottom}.}
	\label{tab:sample_criteria}
	\centering
	\begin{tabular}{cccc}
        \hline\hline
		Selection & Objects qualified & Objects remaining & \textcolor{black}{Significance} \\
		\hline
		\citetalias{Liu19} (broad lines) & 14584 & 14584 & \\
		$\beta_{\mathrm{UV}} < -0.2$, $\beta_{\mathrm{OPT}} > 0$, $\beta_{\mathrm{OPT}} - \beta_{\mathrm{UV}} > 0.5$ & 2778 & 2778 & \textcolor{black}{2$\sigma$} \\
		$\log_{10} L_{5100} - \log_{10} L_{\mathrm{H}\alpha} < 1.4$  & 2384 & 112 & \textcolor{black}{1$\sigma$} \\
		$\log_{10} L_{5100}/L_{\mathrm{MBB}} + 2 \cdot \lambda_{\mathrm{peak}}  > 0.5$ & 3348 & 17 & \textcolor{black}{1$\sigma$} \\
		$\beta_{\mathrm{MBB}} > -3$, 
        \textcolor{black}{$\lambda_{\mathrm{peak}} > 0.5$, }
        $\lambda_{\mathrm{peak}} + 0.1 \cdot \beta_{\mathrm{MBB}} > 0.4$ & 2814 & 15 & \textcolor{black}{1$\sigma$} \\
		$\log_{10} \mathrm{EW}_{\lambda 5007} + \textcolor{black}{2.2} \cdot \log_{10} \mathrm{Balmer\,break\,strength} > \textcolor{black}{1.7}$ & 5880 & 9 & \textcolor{black}{1$\sigma$}  \\
		\hline
        \hline
        \textcolor{black}{Selection for abundance estimation} \\
        \hline
        \textcolor{black}{$\lambda_{\mathrm{peak}} > 3800\,\mathrm{\AA}/(1+z)$, 
        $\lambda_{\mathrm{peak}} < 9200\,\mathrm{\AA}/(1+z)$} & \textcolor{black}{9110} & \textcolor{black}{9110} & \textcolor{black}{1$\sigma$} \\
        \textcolor{black}{all previous cuts} & \textcolor{black}{9} & \textcolor{black}{7} & \textcolor{black}{as stated above} \\
        \hline
	\end{tabular}
\end{table*}
Figure \ref{fig:all_spectra_pop} displays the spectra of the nine selected objects in $f_\lambda$.
For an analogous plot in units of $f_\nu$, see Appendix \ref{app:fnu_sample}.

\subsection{Strictness of the Cuts}\label{sec:strictness}

The selection methods used in this work are meant to find sources which closely match high redshift LRDs in a wide range of properties.
However, LRDs at high redshift are a diverse class of objects.
The comparison to \citetalias{deGraaff25b}, which is a sample constructed with purity in mind, therefore results in a local LRD analogue sample that does not cover the entire range of LRD parameters, but only the more extreme cases.
This suggests that there are many more potential analogue candidates that are not selected here.

Additionally, the parameters based on modified black body fitting ($T$, $\beta_{\mathrm{MBB}}$, $L_{\mathrm{MBB}}$ and $\lambda_{\mathrm{peak}}$) have to be taken with caution if the peak of the fit is not in the wavelength range of the SDSS spectra.
For a fit resulting in a peak at greater wavelength than covered by the spectrum, small changes in the unknown NIR part of the SED can lead to large variations in the fitting parameters.

An analytical problem arises for $\beta_{\mathrm{MBB}}<-4$.
For such values equation (\ref{eq:mbb_fnu}) diverges for high $\lambda$.
This can be seen when looking at the Rayleigh-Jeans limit of the Planck function:
\begin{equation}
	B_{\lambda} \sim \lambda^{-4}
\end{equation}
Multiplying this by the modifying term gives
\begin{equation}
	f_{\mathrm{MBB}}  \sim \lambda^{-4} \cdot \lambda^{-\beta_{\mathrm{MBB}}}
\end{equation}
Inserting any value $<-4$ for $\beta_{\mathrm{MBB}}$ results in 
\begin{equation}
	\lim_{\lambda \to \infty} f_{\mathrm{MBB}} = \infty\,\mathrm{.}
\end{equation}
 This means that no peak wavelength can be determined.
 However, as $\beta_{\mathrm{MBB}}<-4$ is not in the range which is interesting for LRDs, this is not problematic and such objects are excluded from Figures which display $\lambda_{\mathrm{peak}}$.
 
 In addition, for very blue sources, the peak shifts to very short wavelengths which can be located outside of the spectral range, making the fitting parameters very uncertain.
 Again, these objects are not interesting as LRD candidates as they would necessarily have very blue optical continua.
 However, as their best-fit values in our model have high uncertainties, they are also omitted from the Figures if $\lambda_{\mathrm{peak}} < 0.1\,\mathrm{\mu m}$.

 Lastly, if the SED is rather flat in the range used for the MBB fitting, the recovered $\lambda_{\mathrm{peak}}$ becomes very uncertain and objects might miss the selection cuts.
 In summary, only for objects with a $\lambda_{\mathrm{peak}}$ confidently within the detector range of the SDSS spectrograph ($3800-9200\,\mathrm{\AA}$), we confidently can determine if such objects are LRD analogues or not.
 Due to these limitations, some LRD analogues might be missed with this procedure, if they do not fulfil  $\lambda_{\mathrm{peak}} > 3800\,\mathrm{\AA}/(1+z)$ and $\lambda_{\mathrm{peak}} < 9200\,\mathrm{\AA}/(1+z)$.
 9110 objects are within these cuts at the 1$\sigma$ confidence and only seven of the nine local analogues.
 Therefore, accounting for this we correct the fraction of local LRD analogues up to $0.08\%$.
 One further caveat of our search is that in \citetalias{deGraaff25b} some LRDs have a very ``cold'' MBB, meaning their $\lambda_{\mathrm{peak}}$ is very high and outside the SDSS spectrograph for the \citetalias{Liu19} sample.
 Depending on the redshift of the local objects ($0<z<0.35$), the fraction of missed objects due to this limitation varies between $9\%$ to $51\%$.
 This means the fraction of local LRD analogues could potentially be as high as $0.16\%$.
 This could be alleviated in the future by implementing SPHEREx data for the entire sample, which is not yet possible due to challenges in retrieving data.

\section{Results}\label{sec:sample_analysis}
\begin{figure*}
    \includegraphics[width=17cm]{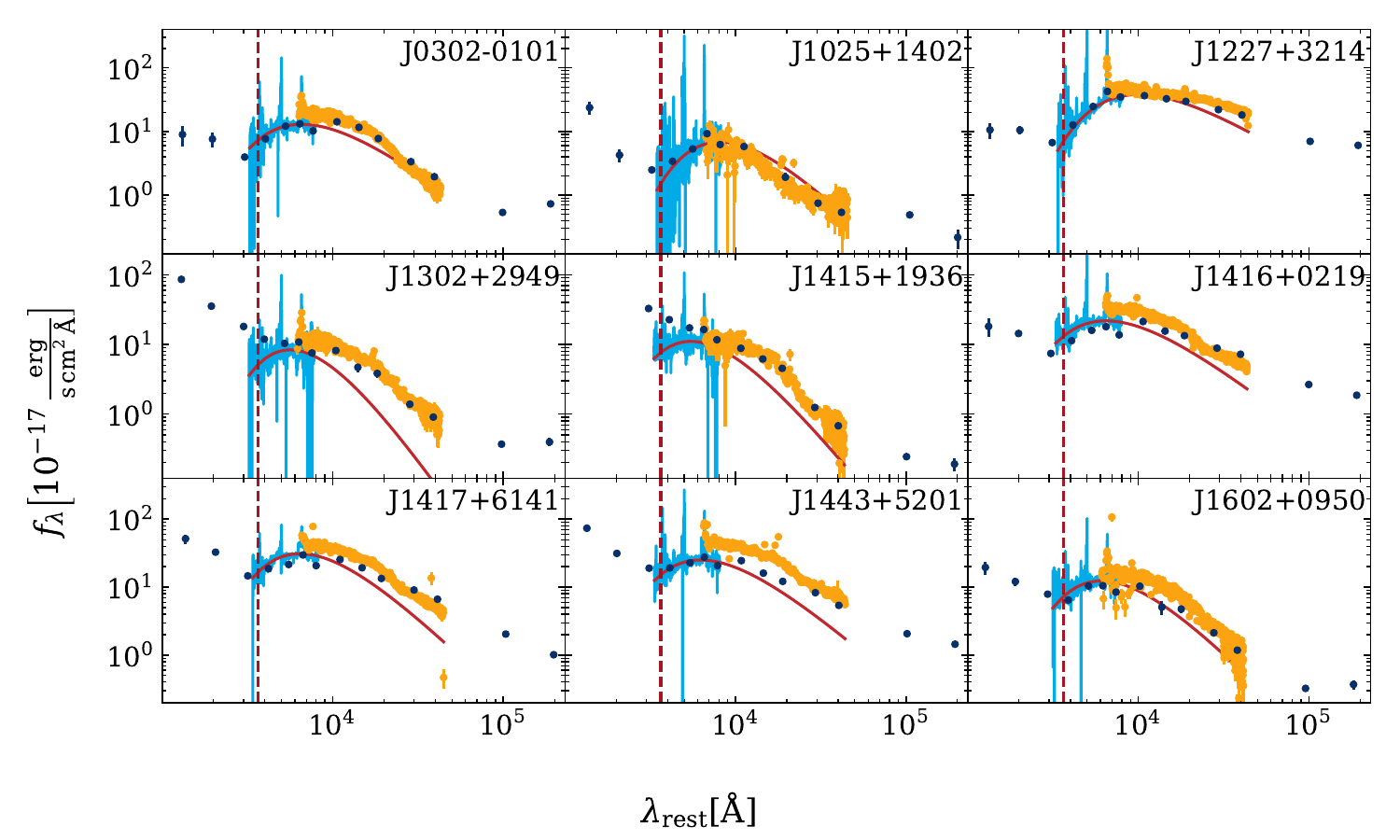}
	\caption{Spectra in $f_\lambda$ of our 9 local LRD analogues, using SDSS (light blue), photometric data (navy blue), and SPHEREx spectrophotometry (yellow). The red dashed line indicates the Balmer limit, \textcolor{black}{while the solid line shows the fitted modified black-body}. For some sources, the SPHEREx spectra are offset from SDSS. This could be due to contamination of SPHEREx observations from extended hosts.}
	\label{fig:all_spectra_pop}
\end{figure*}

\begin{figure}
	\resizebox{\hsize}{!}{\includegraphics{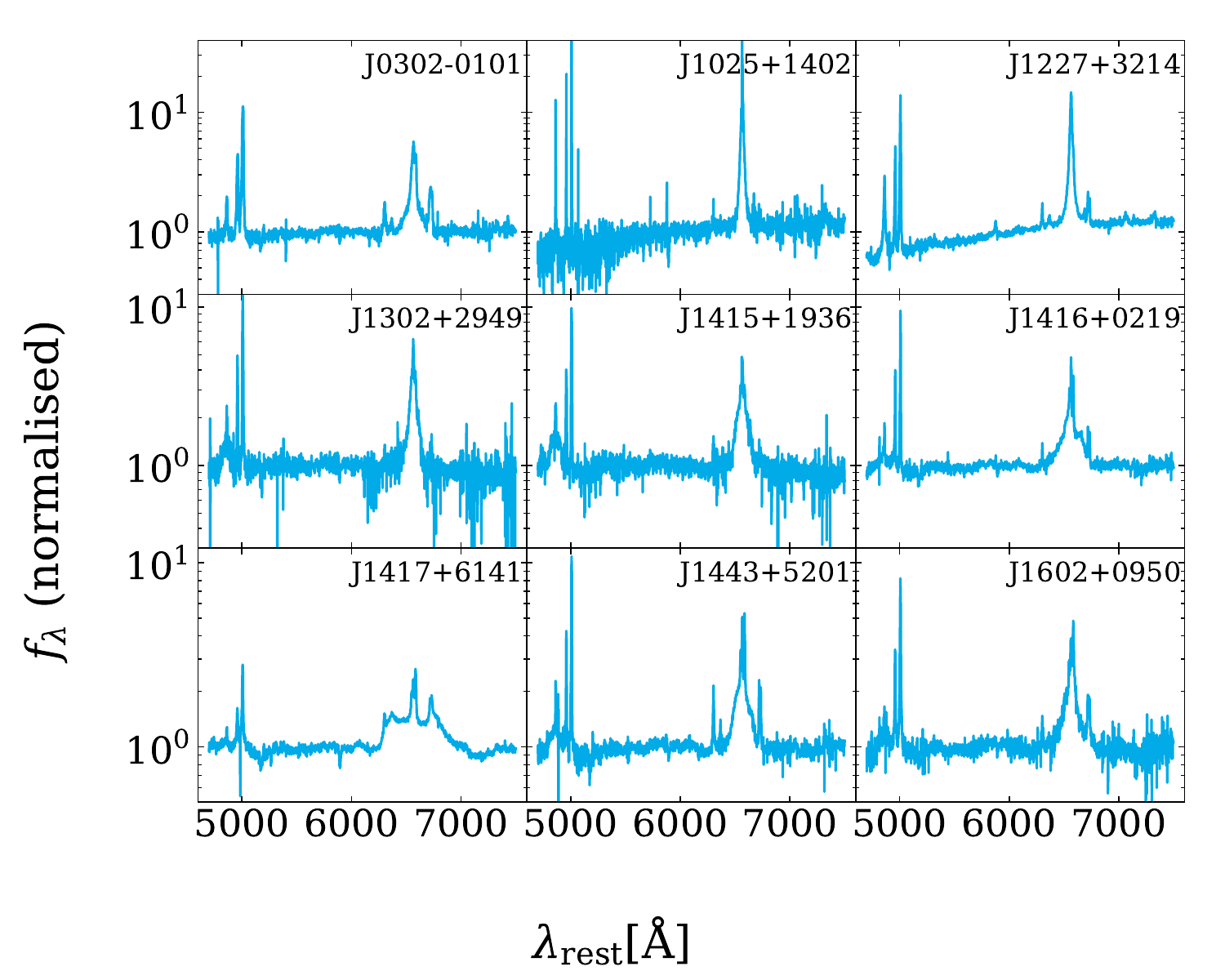}}
	\caption{\textcolor{black}{Normalised} spectra in $f_\lambda$ of all 9 local analogues from SDSS focused on the region between H$\beta$ and H$\alpha$.}
	\label{fig:all_spectra_pop_Hab}
\end{figure}

The sample selection yields 9 local LRD analogues.
Figure \ref{fig:all_spectra_pop} shows their rest-frame optical spectra with SPHEREx spectrophotometry in the NIR and photometry in the optical and IR.
In Figure \ref{fig:all_spectra_pop_Hab} the H$\alpha$ and H$\beta$ spectral regions are shown more closely.
Finally, another property that is of high interest in LRDs is variability, since it
could help to distinguish between models.
ZTF light curves are gathered for the $g$-band and presented in Appendix Figure \ref{app:ztf_sample}.

\subsection{General properties}
One of the objects we selected, J1025+1402, had previously been identified as an LRD analogue \citep[The Egg/Lord of LRDs;][]{LinX25b,Ji25b}.
The re-selection of this object is a validation of 
the selection constructed in this work.

In five of the nine objects, including the Egg, the MBB fit obtained from the rest-frame optical is also in very good agreement with the rest-frame NIR $-$ which was not used in the selection (J0302$-$0101, J1227+3214, J1416+0219, and J1602+0950). Two other objects show mild exceeses in the NIR compared to the MBB fit (J1417+6141 and J1443+5201), which would have shifted $\lambda_\text{peak}$ to slightly longer wavelength if the NIR data had been included, but ultimately  would not have affected their selection. 

The last two objects (J1302+2949 and J1415+1936) are more peculiar: in addition to very large excesses in the rest-frame NIR compared to the VIS-derived MBB fit, both show strong disagreement between the optical spectra and the optical photometry. Both objects also show the clearest signs of variability in their ZTF light curves. This raises the tantalising possibility that they may have transitioned ``out'' of an LRD-like phase.

In addition to their LRD-like SEDs, all nine sources show large Balmer decrements ($3<\text{BD}<11$) and clear broad H$\alpha$ lines. Their FWHM range from $\sim1300-5900$ km s${}^{-1}$, with the exception of J1417+6141 which displays an extremely unusual and puzzling ``square'' H$\alpha$ line morphology. 
The majority of the nine objects are mostly compact or show a strong compact core (see also \ref{app:photometry}).
Only one of the objects, J1443+5201, is detected in the radio.

\subsection{Individual Objects}

In this Section we outline the properties of the nine analogues object-by-object.

\subsubsection{J0302$-$0101}
J0302$-$0101 continues to agree with the MBB fit even in the NIR observations from SPHEREx and 2MASS and W1 and W2.
In the range of W3 and W4, the SED seems to flatten and turn over.
The spectrum emission shows lines matching [O~{\small{I}]}$\lambda 6302$ and the [S~{\small{II}}]$\lambda \lambda 6716,6731$ doublet.
The [N~{\small{II}}]$\lambda \lambda 6548,6583$ doublet is weak and blended with H$\alpha$, but also present.
A broad component is only visible in H$\alpha$, but not in H$\beta$.
The object shows no variability in ZTF photometry.
In optical photometry (\ref{app:photometry}) it shows a extended host, but also has a compact core.


\subsubsection{J1025+1402}
Part of the selection is J1025+1402, which is a previously-known local LRD analogue \citep{LinX25b,Ji25b}, originally was presented by \citet{Izotov08} as a metal-poor emission line galaxy with an AGN.
J1025+1402 has an absorption feature in H$\alpha$ and agrees with the MBB in the NIR.
J1025+1402 shows signs of a Na D absorption feature at $5896\,\mathrm{\AA}$, accompanied by an emission line blueward of the absorption, which matches the expected location of He~{\small{I}} at $5786\,\mathrm{\AA}$ $-$ a line commonly found in LRDs \citep{Matthee26}.
It is the only source in the selection that does not show an [N~{\small{II}}] emission line, and [O~{\small{I}}]$\lambda 6302$ is only detected tentatively. 
The broad emission is prominent in H$\alpha$, but very weak in H$\beta$.
It shows no variability in ZTF photometry and has a compact morphology.

\subsubsection{J1227+3214}
J1227+3214 is consistent with the MBB into the NIR, but starts to deviate at longer wavelengths, where it is slightly redder. This may indicate the presence of hot dust in the AGN accretion structure. 
It shows a clear [S~{\small{II}}] doublet and [O~{\small{I}}]$\lambda 6302$ emission.
The H$\alpha$ line shows a slight red shoulder, which corresponds to the [N~{\small{II}}]$\lambda 6583$ line.
J1227+3214 also displays Na D absorption as well as the nearby He~{\small{I}} line.
Variability in ZTF is tentatively present, but a more sophisticated analysis would need to confirm this.
The morphology is compact in the optical.

\subsubsection{J1302+2949}
J1302+2949 was first discovered by \citet{Barbieri72} as a faint blue source and later as a UV-excess quasar \citep{Moreau95}.
It shows a concerning mismatch between the optical photometry and the spectrum, which is redder.
When looking solely at the photometry, there is no sign of a v-shape.
Therefore, its classification as a local LRD analogue should be taken with caution. Alternatively, the mismatch could be explained by variability: the ZTF light-curve of J1302+2949 shows variability that spans more than a magnitude, reaching a regime where it could be considered a changing-look AGN. The object be therefore be transition out of an ``LRD phase'' captured by the spectroscopy.

The MBB fit in the NIR is bluer than the SPHEREx and 2MASS photometry; however, we note that SPHEREx spectrum was taken at a significantly more recent time.
Nevertheless, the SED still flattens similarly to e.g.~J1025+1402.
The Na D absorption as well as the [O~{\small{I}}]$\lambda 6302$ line seen in the other sources are both very weak if even present at all in J1302+2949.
The [N~{\small{II}}] and [S~{\small{II}}] doublets are also very weak.
The morphology shows a slightly extended host.

\subsubsection{J1415+1936}
J1415+1936 shows the same concerning mismatch between photometry and spectrum as J1302+2949. Its light-curve shows clear signs of variability over about 0.5 mag, lending itself to a similar interpretation as J1302+2949.

The MBB is slightly bluer than the NIR observations, but the SED still flattens.
The [O~{\small{I}}], [N~{\small{II}}] and [S~{\small{II}}] emission lines are all very weak and a Na D absorption line is only tentatively detected.
Morphologically, it shows a compact core accompanied with a extended host.

\subsubsection{J1416+0219}
J1416+0219 shows complex broad Balmer line morphology. 
A broad drop in the spectrum, which may be interpreted as absorption, is visible redwards of the line centre for both H$\alpha$ and H$\beta$.
J1416+0219 is part of the sample of \citet{Strateva03}, where it is classified as an double peaked emitter \citep[DPE; see e.g. ][]{Ward25}. 
\textcolor{black}{Why a DPE should display a v-shaped SED however, is not clear.}

The NIR observations agree well with the MBB fit, however the SED only flattens slightly compared to the optical.
Both the [N~{\small{II}}] and [S~{\small{II}}] doublets are clearly present and the [O~{\small{I}}] line is also seen.
Na D absorption is detected as well as the He~{\small{I}} emission line nearby.
ZTF photometry shows no signs of variability and in morphology a compact core is surrounded by an extended galaxy.

\begin{table}
	\caption{Basic properties of the selected local LRD analogues for the criteria worked out in Section \ref{sec:sample_construction}.}
	\label{tab:analogues_props_basic}
	\centering
	\begin{tabular}{cccc}
		\hline\hline   
		SDSS Name & RA & DEC & $z$ \\
		\hline
		J0302$-$0101 & 45.505091 & $-$1.0216328 & 0.166488 \\
		J1025+1402 & 156.37622 & 14.035377 & 0.10064 \\
		J1227+3214 & 186.95475 & 32.249702 & 0.136826 \\
		J1302+2949 & 195.7267 & 29.820841 & 0.183646 \\
		J1415+1936 & 213.92211 & 19.615524 & 0.149458 \\
		J1416+0219 & 214.05569 & 2.3188421 & 0.158219 \\
		J1417+6141 & 214.42896 & 61.697924 & 0.11887 \\
		J1443+5201 & 220.7615 & 52.027006 & 0.14121 \\
		J1602+0950 & 240.72893 & 9.8360933 & 0.209512 \\
		\hline
	\end{tabular}
\end{table}

\subsubsection{J1417+6141}
J1417+6141 shows extremely complex Balmer emission as well. The feature in its spectrum could be described as a plateau, with elevated emission from approximately $6250\,\mathrm{\AA}$ to $7000\,\mathrm{\AA}$.
This extent means that the plateau has a distinctly ``square'' look and appears to be roughly centred around H$\alpha$.
This object was analysed individually in \citet{WangT05} as a DPE.

The MBB fit from the rest-frame optical does not agree well with the rest-frame NIR photometry: a joint fit of the two would yield a redder $\lambda_\text{peak}$. The SED flattens only slightly towards the NIR. In this object, the SPHEREx spectrophotometry is offset, probably due to contamination. 
At SPHEREx's much higher fibre size, the NIR continuum around $1\,\mu \mathrm{m}$ rest-frame probably picks up stellar emission from the extended host galaxy.
The object shows Na D absorption, but no He~{\small{I}} emission line. [N~{\small{II}}], [S~{\small{II}}] and [O~{\small{I}}] are all detected.
The light-curve of J1417+6141 shows no variability.

\subsubsection{J1443+5201}
J1443+5201 is a long-known radio source \citep{Edge59}.
However, the radio properties of LRDs are still very uncertain and current measurements can only provide upper limits that could be consistent with radio-quiet AGNs.
As in the previous object, The MBB fit and NIR observations disagree and the SED is still slightly red. Also as in the previous object, the SPHEREx spectrum is likely contaminated by the extended host.

H$\alpha$ shows signs of slightly asymmetric shoulders, which could indicate multiple broad components. 
[N~{\small{II}}], [S~{\small{II}}] and [O~{\small{I}}] are detected strongly.
The Na D absorption is accompanied by tentative He~{\small{I}}.
Additionally, a [O~{\small{I}}]$\lambda 6365$ line is seen.
ZTF photometry shows tentative signs of variability.

\subsubsection{J1602+0950}
For J1602+0950 the MBB fit agrees with 2MASS and \textit{WISE} W1 and W2 photometry, but again the SPHEREx spectrophotometry is offset, probably due to the extended host.
Alternatively, this could be due to the source varying between the different observations, as  the ZTF light-curve shows clear signs of variability despite very poor temporal sampling.
Nevertheless, all NIR observations show a clear flattening of the SED.
The [N~{\small{II}}], [S~{\small{II}}] and [O~{\small{I}}] emission lines as well as weak Na D absorption is detected, but no He~{\small{I}} line.

\section{Discussion}\label{sec:discussion}
\subsection{Local LRD Analogues}
The nine objects in the main sample show observational properties that mimic LRDs in various ways.
This does not necessarily mean that they are the same physical objects.
Still, it is valuable to see which kind of objects can look like LRDs.
For some of the objects, a first classification attempt is rather easy, as they have been studied before.

Two of our analogues (J1416+0219 and J1417+6141) display complex H$\alpha$ emission line morphologies which have both been interpreted as double peaked emitters (DPEs) in previous literature.
Therefore, their classification as local LRD analogues should be taken with caution.
While LRDs have symmetric exponential profiles, DPEs show highly asymmetric profiles.
Due to their double-peaked nature, the emission line fluxes of DPEs can be very high $-$ helping them
satisfy our selection cut on $L_{\mathrm{H}\alpha}/L_{5100}$.
However, they still fulfil all other selection cuts as well, especially those relating to their SEDs: J1416+0219 in particular is excellently fit by a MBB.
As a potential source of contamination, DPEs can be easily excluded with a simple inspection of the lines.

\begin{figure*}
    \includegraphics[width=17cm]{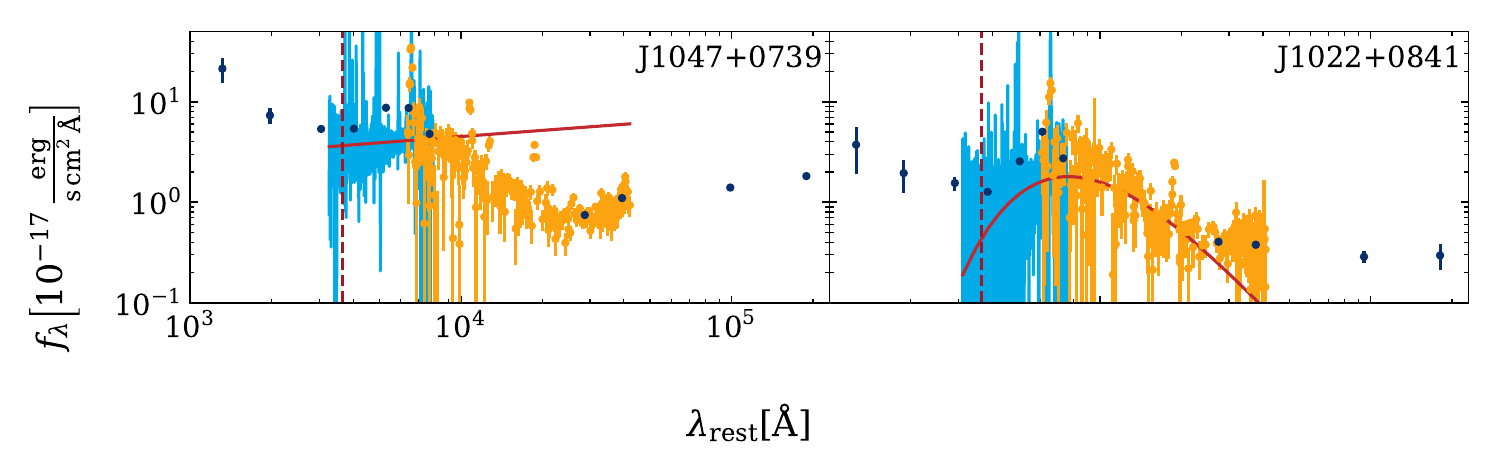}
	\caption{SDSS Spectra of the two local LRD analogues from \citet{LinX25b} that are not selected as local LRD analogues in this study. \textcolor{black}{Symbols and colours are the same as described in Fig. \ref{fig:all_spectra_pop}.}}
	\label{fig:all_spectra_pop_failed_llrds}
\end{figure*}

A straightforward answer to whether the sample presented here shares the physical nature of LRDs can not be given.
For one, the physical processes at work within LRDs remain elusive $-$ in that regard we hope that this sample could give inspiration for interpretation.
Secondly, an in-depth study of each of the selected sources would be necessary to clarify the physical components of these objects.
Follow-up observations would be needed to confirm the entire v-shape region spectroscopically.
This is especially true for J1302+2949 and J1415+1936, which show a deviation of photometry and the SDSS spectra in wavelengths shorter than the Balmer break $-$ both objects are also clearly variable.

There are some properties often reported for high redshift LRDs that are not used in this work.
This includes the NIR behaviour, the Balmer decrement as well as lack of variability.
Those are not used in the selection as there is either no clear observational consensus or sufficiently substantial sample for these quantities in the literature yet. 
Depending on future developments in high redshift LRD observations, corresponding criteria could be implemented in a future study.

When considering the argument from \citet{Matthee26} and \citet{Billand26} that current high-$z$ LRD selections are significantly biased to only recover the extreme end of a broader population, we acknowledge that (by design) the nine analogues presented here are also only representative of that extreme. 
When relaxing the selection criteria, one finds many more objects that qualify as LRD analogues (see Fig,~\ref{fig:population_map}). 
The viability and usefulness of such an attempt is unclear, since the more the criteria are relaxed the closer the selected objects come to the general AGN sample and therefore more contamination is expected. In other words, LRD analogues may represent one extreme of a continuum of AGN properties, making the ``edge'' of criteria difficult to define. 
Ultimately, the purpose of our low-$z$ search for LRD analogues is motivated by the hope of finding local, easily accessible examples of the physical processes giving rise to LRDs's most peculiar features (v-shape and MBB-like SED, broad lines). 
Therefore, the selection we implemented is rather strict, and all cuts demanded at least a $1\sigma$ confidence.
If this constraint is lifted, many more sources are selected (60 sources). 
For those, the LRD analogue classification is more questionable and would require a case-by-case analysis, but they provide an excellent candidate sample, which could be investigated with follow-up observations.

\subsection{Missed Objects}\label{sec:missed_objects}
The sample selection is limited by the available wavelength coverage and depth.
This can be illustrated best with the two local LRD analogues from \citet{LinX25b} that are not part of the final sample, but are part of the \citetalias{Liu19} catalogue (J1047+0739 and J1022+0841; Figure \ref{fig:all_spectra_pop_failed_llrds}).

For J1047+0739, the issue is a very flat optical continuum without a peak within the SDSS spectrum.
\textcolor{black}{The result is a $\beta_{\text{MBB}} < -4$ (Table \ref{tab:analogues_props}), for which the modified black-body diverges at high wavelength and $\lambda_{\text{peak}}$ and $L_{\text{MBB}}$ are infinite and lose physical meaning (see Sec.~\ref{sec:strictness}).}
With the addition of SPHEREx data or follow-up observations as in \citet{LinX25b}, this issue can be alleviated.
Due to technical limitations in retrieving SPHEREx data from IRSA, this is not yet possible for the entire sample, but developments of an improved pipeline may soon make it feasible \citep{Davies26}.
Additionally, contamination from nearby sources in SPHEREx and the fact that some sources may be extended in the NIR limits a trivial implementation.

The other object we missed, J1022+0841, shows a redder optical and the MBB fitting does find a solution that is confirmed by the SPHEREx spectrum, but the rest-frame optical spectrum has an SNR which is too low. 
\textcolor{black}{This is reflected in the high uncertainties of the fitting parameters and the parameters calculated from those as $L_{\text{5100}}/L_{\text{MBB}}$ (see Table \ref{tab:analogues_props}), which is the only cut J1022+0841 misses due to the required 1$\sigma$ significance.}
The uncertainties put it out of the confidence range, while the values themselves fulfil the criteria.
Unfortunately, this issue can only be resolved with better quality observations.

\section{Conclusions}\label{sec:summary}

We selected LRD analogues at $0<z<0.35$ starting from the unbiased parent catalogue of \citetalias{Liu19}, which contains all objects with broad H$\alpha$ and H$\beta$ emission lines (both galaxies and AGN). 
We enforce the selection of a v-shaped SED and broad Balmer lines to resemble the $2.3<z<9.3$ LRD sample of \citetalias{deGraaff25b}. 
A criterium of compactness is not used, as it may be incidental low redshift LRDs.

We use SDSS-II and, if available, eBOSS spectra. The SEDs are complemented by SDSS, GALEX, 2MASS, \textit{WISE}, SPHEREx, Legacy Survey and ZTF (spectro-)photometry.

We define SEDs as having a v-shape based on their rest-frame UV and rest-frame optical slopes, and further fit the SED with a modified black body (MBB) model. The best-fit parameters of the MBB fit are used to identify objects with SEDs similar to high-$z$ LRDs.
We find that 19\% of all objects with broad lines display a v-shaped SED, while 0.08\% are well-fit by LRD-like MBBs. Among low-$z$ objects which match the H$\alpha$ and continuum luminosities of high-$z$ LRDs, 5\% have a v-shaped SED and 0.37\% have an LRD-like MBB SED.

Ultimately, nine objects are selected as local LRD analogues, one of which being a previously-identified analogue nicknamed ``The Egg''.
The other eight objects show a variety of odd properties which may help explain their unusual SEDs. One object, J0302$-$0101, shows all the signs expected of an LRD, even though it appears imbedded into a host galaxy (which contributes very little to its rest-frame optical spectrum). Another object, J1227+3214, has an ambiguous nature, since its rest-frame IR emission resembles an unattenuated bright quasar but its rest-frame optical and UV SED is extremely peculiar. Two objects are known double-peaked AGN, two show very strong variability, and two display significant contamination by large host galaxies.

In conclusion, our work highlights that LRD-like objects are extremely rare at $z<0.35$, even when exclusion criteria on compactness are not enforced. This is consistent with literature which has found a sharp decline in the number density of LRDs with cosmic time \citep{Ma25}.


Several pathways for expanding this work in the future exist. 
With more time and more efficient data access, SPHEREx spectrophotometry can be gathered for all sources. 
This would allow for a much better constraint on the modified black body fits for our sources, especially when the peak of the emission is redwards of the SDSS/eBOSS spectroscopic range.
Follow-up observations and individual in depth-analyses of the nine selected analogues could confirm or reject a physical connection to LRDs. 
Similarly, follow-up of candidate sources that narrowly fail the $1\sigma$ exclusion criteria could lead to the discovery of further local analogues.
As new LRD observations continue to be published, new insights into the LRD population can be implemented into the selection.

\begin{acknowledgements}
\textcolor{black}{We thank the anonymous referee for constructive suggestions, which helped to improve the quality of this manuscript. LS acknowledges support from the German Space Agency (DLR) through Verbundforschung project ID 50OR2512}. SEIB is supported by the Deutsche Forschungsgemeinschaft (DFG) under Emmy Noether grant number BO 5771/1-1. The authors acknowledge productive discussions with Frederick Davies.

This publication makes use of data products from the Spectro-Photometer for the History
of the Universe, Epoch of Reionization and Ices Explorer (SPHEREx), which
is a joint project of the Jet Propulsion Laboratory and the California Institute
of Technology, and is funded by the National Aeronautics and Space Administration.
This publication makes use of data products
from the Wide-field Infrared Survey Explorer, which is a joint project of the University of California, Los Angeles, and the Jet Propulsion Laboratory/California
Institute of Technology, funded by the National Aeronautics and Space Administration.
Based on observations obtained with the Samuel Oschin Telescope 48-inch and the 60-inch Telescope at the Palomar
Observatory as part of the Zwicky Transient Facility project. ZTF is supported by the National Science Foundation under Grants
No. AST-1440341 and AST-2034437 and a collaboration including current partners Caltech, IPAC, the Oskar Klein Center at
Stockholm University, the University of Maryland, University of California, Berkeley , the University of Wisconsin at Milwaukee,
University of Warwick, Ruhr University, Cornell University, Northwestern University and Drexel University. Operations are
conducted by COO, IPAC, and UW.
This publication makes use of data products from the Two Micron All Sky Survey, which is a joint project of the University of Massachusetts and the Infrared Processing and Analysis Center/California Institute of Technology, funded by the National Aeronautics and Space Administration and the National Science Foundation.
This research is based on observations made with the Galaxy Evolution Explorer, obtained from the MAST data archive at the Space Telescope Science Institute, which is operated by the Association of Universities for Research in Astronomy, Inc., under NASA contract NAS 5–26555.
Funding for the SDSS and SDSS-II has been provided by the Alfred P. Sloan Foundation, the Participating Institutions, the National Science Foundation, the U.S. Department of Energy, the National Aeronautics and Space Administration, the Japanese Monbukagakusho, the Max Planck Society, and the Higher Education Funding Council for England. The SDSS Web Site is http://www.sdss.org/.

The SDSS is managed by the Astrophysical Research Consortium for the Participating Institutions. The Participating Institutions are the American Museum of Natural History, Astrophysical Institute Potsdam, University of Basel, University of Cambridge, Case Western Reserve University, University of Chicago, Drexel University, Fermilab, the Institute for Advanced Study, the Japan Participation Group, Johns Hopkins University, the Joint Institute for Nuclear Astrophysics, the Kavli Institute for Particle Astrophysics and Cosmology, the Korean Scientist Group, the Chinese Academy of Sciences (LAMOST), Los Alamos National Laboratory, the Max-Planck-Institute for Astronomy (MPIA), the Max-Planck-Institute for Astrophysics (MPA), New Mexico State University, Ohio State University, University of Pittsburgh, University of Portsmouth, Princeton University, the United States Naval Observatory, and the University of Washington.
Funding for the Sloan Digital Sky 
Survey IV has been provided by the 
Alfred P. Sloan Foundation, the U.S. 
Department of Energy Office of 
Science, and the Participating 
Institutions. 

SDSS-IV acknowledges support and 
resources from the Center for High 
Performance Computing  at the 
University of Utah. The SDSS 
website is www.sdss4.org.

SDSS-IV is managed by the 
Astrophysical Research Consortium 
for the Participating Institutions 
of the SDSS Collaboration including 
the Brazilian Participation Group, 
the Carnegie Institution for Science, 
Carnegie Mellon University, Center for 
Astrophysics | Harvard \& 
Smithsonian, the Chilean Participation 
Group, the French Participation Group, 
Instituto de Astrof\'isica de 
Canarias, The Johns Hopkins 
University, Kavli Institute for the 
Physics and Mathematics of the 
Universe (IPMU) / University of 
Tokyo, the Korean Participation Group, 
Lawrence Berkeley National Laboratory, 
Leibniz Institut f\"ur Astrophysik 
Potsdam (AIP),  Max-Planck-Institut 
f\"ur Astronomie (MPIA Heidelberg), 
Max-Planck-Institut f\"ur 
Astrophysik (MPA Garching), 
Max-Planck-Institut f\"ur 
Extraterrestrische Physik (MPE), 
National Astronomical Observatories of 
China, New Mexico State University, 
New York University, University of 
Notre Dame, Observat\'ario 
Nacional / MCTI, The Ohio State 
University, Pennsylvania State 
University, Shanghai 
Astronomical Observatory, United 
Kingdom Participation Group, 
Universidad Nacional Aut\'onoma 
de M\'exico, University of Arizona, 
University of Colorado Boulder, 
University of Oxford, University of 
Portsmouth, University of Utah, 
University of Virginia, University 
of Washington, University of 
Wisconsin, Vanderbilt University, 
and Yale University.

The Legacy Surveys consist of three individual and complementary projects: the Dark Energy Camera Legacy Survey (DECaLS; Proposal ID \#2014B-0404; PIs: David Schlegel and Arjun Dey), the Beijing-Arizona Sky Survey (BASS; NOAO Prop. ID \#2015A-0801; PIs: Zhou Xu and Xiaohui Fan), and the Mayall z-band Legacy Survey (MzLS; Prop. ID \#2016A-0453; PI: Arjun Dey). DECaLS, BASS and MzLS together include data obtained, respectively, at the Blanco telescope, Cerro Tololo Inter-American Observatory, NSF’s NOIRLab; the Bok telescope, Steward Observatory, University of Arizona; and the Mayall telescope, Kitt Peak National Observatory, NOIRLab. Pipeline processing and analyses of the data were supported by NOIRLab and the Lawrence Berkeley National Laboratory (LBNL). The Legacy Surveys project is honored to be permitted to conduct astronomical research on Iolkam Du’ag (Kitt Peak), a mountain with particular significance to the Tohono O’odham Nation.

NOIRLab is operated by the Association of Universities for Research in Astronomy (AURA) under a cooperative agreement with the National Science Foundation. LBNL is managed by the Regents of the University of California under contract to the U.S. Department of Energy.

This project used data obtained with the Dark Energy Camera (DECam), which was constructed by the Dark Energy Survey (DES) collaboration. Funding for the DES Projects has been provided by the U.S. Department of Energy, the U.S. National Science Foundation, the Ministry of Science and Education of Spain, the Science and Technology Facilities Council of the United Kingdom, the Higher Education Funding Council for England, the National Center for Supercomputing Applications at the University of Illinois at Urbana-Champaign, the Kavli Institute of Cosmological Physics at the University of Chicago, Center for Cosmology and Astro-Particle Physics at the Ohio State University, the Mitchell Institute for Fundamental Physics and Astronomy at Texas A\&M University, Financiadora de Estudos e Projetos, Fundacao Carlos Chagas Filho de Amparo, Financiadora de Estudos e Projetos, Fundacao Carlos Chagas Filho de Amparo a Pesquisa do Estado do Rio de Janeiro, Conselho Nacional de Desenvolvimento Cientifico e Tecnologico and the Ministerio da Ciencia, Tecnologia e Inovacao, the Deutsche Forschungsgemeinschaft and the Collaborating Institutions in the Dark Energy Survey. The Collaborating Institutions are Argonne National Laboratory, the University of California at Santa Cruz, the University of Cambridge, Centro de Investigaciones Energeticas, Medioambientales y Tecnologicas-Madrid, the University of Chicago, University College London, the DES-Brazil Consortium, the University of Edinburgh, the Eidgenossische Technische Hochschule (ETH) Zurich, Fermi National Accelerator Laboratory, the University of Illinois at Urbana-Champaign, the Institut de Ciencies de l’Espai (IEEC/CSIC), the Institut de Fisica d’Altes Energies, Lawrence Berkeley National Laboratory, the Ludwig Maximilians Universitat Munchen and the associated Excellence Cluster Universe, the University of Michigan, NSF’s NOIRLab, the University of Nottingham, the Ohio State University, the University of Pennsylvania, the University of Portsmouth, SLAC National Accelerator Laboratory, Stanford University, the University of Sussex, and Texas A\&M University.

BASS is a key project of the Telescope Access Program (TAP), which has been funded by the National Astronomical Observatories of China, the Chinese Academy of Sciences (the Strategic Priority Research Program “The Emergence of Cosmological Structures” Grant \# XDB09000000), and the Special Fund for Astronomy from the Ministry of Finance. The BASS is also supported by the External Cooperation Program of Chinese Academy of Sciences (Grant \# 114A11KYSB20160057), and Chinese National Natural Science Foundation (Grant \# 12120101003, \# 11433005).

The Legacy Survey team makes use of data products from the Near-Earth Object Wide-field Infrared Survey Explorer (NEOWISE), which is a project of the Jet Propulsion Laboratory/California Institute of Technology. NEOWISE is funded by the National Aeronautics and Space Administration.

The Legacy Surveys imaging of the DESI footprint is supported by the Director, Office of Science, Office of High Energy Physics of the U.S. Department of Energy under Contract No. DE-AC02-05CH1123, by the National Energy Research Scientific Computing Center, a DOE Office of Science User Facility under the same contract; and by the U.S. National Science Foundation, Division of Astronomical Sciences under Contract No. AST-0950945 to NOAO.

This research has made
use of ``Aladin sky atlas'' developed at CDS, Strasbourg Observatory, France.
This research made use of \texttt{NumPy} \citep{numpy}, \texttt{SciPy} \citep{scipy}, \texttt{Astropy} \citep{astropy:2013, astropy:2018, astropy:2022}, \texttt{Sculptor} \citep{sculptor} and \texttt{Matplotlib} \citep{matplotlib}.
\end{acknowledgements}


\bibliographystyle{aa}
\bibliography{ref}

\begin{appendix}

\section{Sample Selection}\label{app:sample_selection}
\subsection{The Nine in $f_\nu$}\label{app:fnu_sample}
\begin{figure}
	\resizebox{\hsize}{!}{\includegraphics[width=\textwidth]{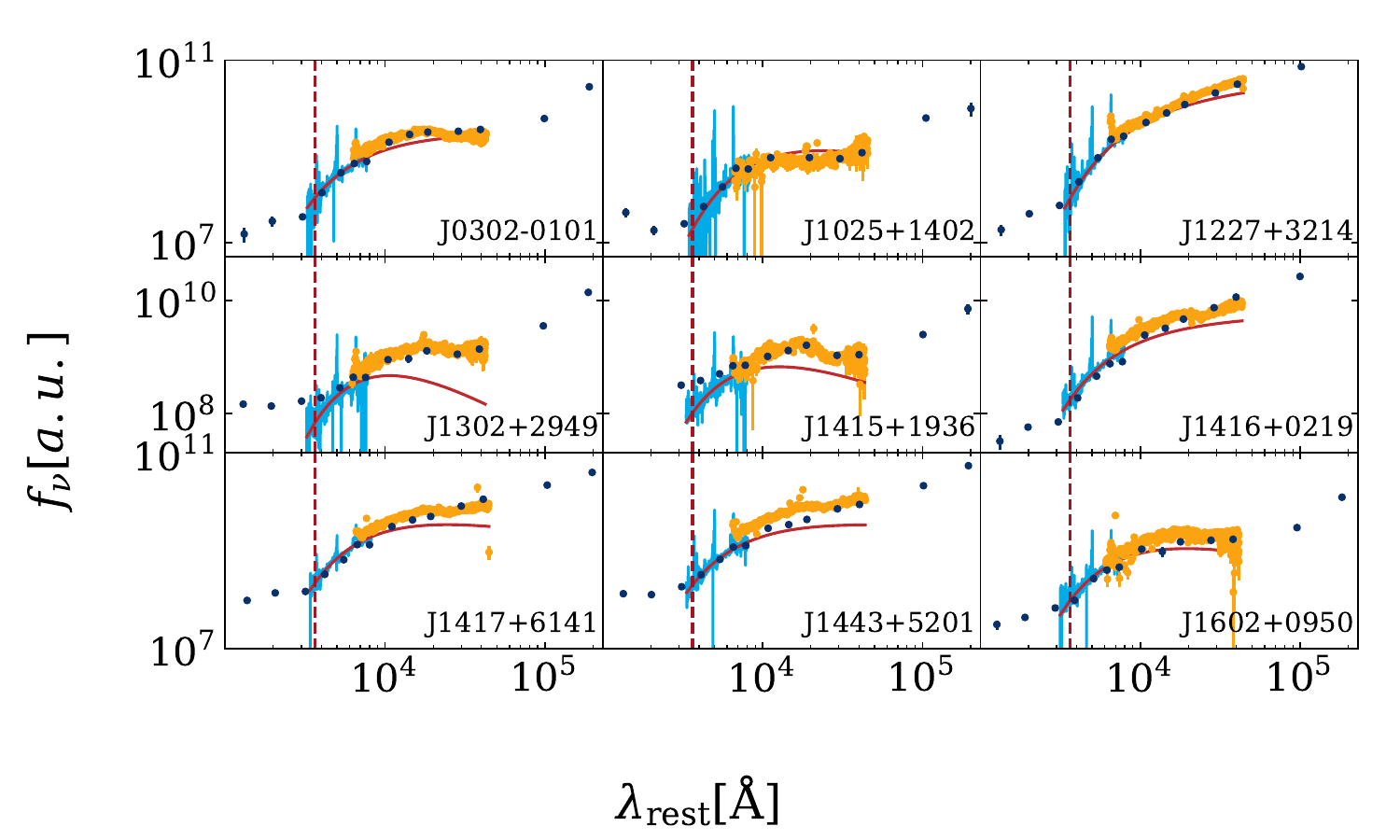}}
	\caption{Spectra in $f_\nu$ of all local analogues from SDSS (light blue) with photometric data (navy blue) and SPHEREx spectrophotometry (yellow). The red dashed line indicates the Balmer limit \textcolor{black}{while the solid line shows the fitted modified black-body}.}
	\label{fig:all_spectra_pop_fnu}
\end{figure}

\begin{figure}
	\resizebox{\hsize}{!}{\includegraphics{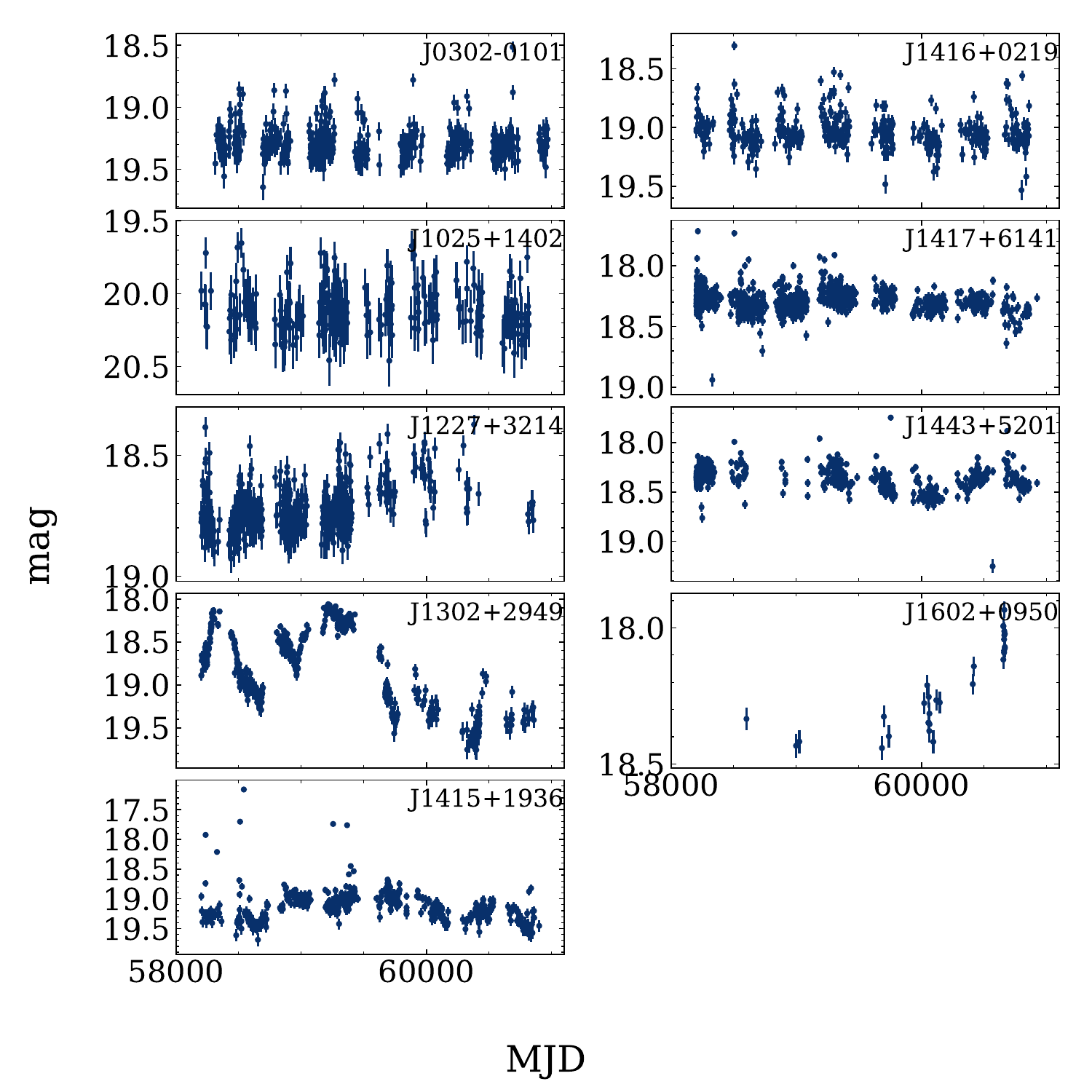}}
	\caption{ZTF $g$-band light curves for the selected analogue sample.}
	\label{app:ztf_sample}
\end{figure}

\begin{figure}
	\resizebox{\hsize}{!}{\includegraphics{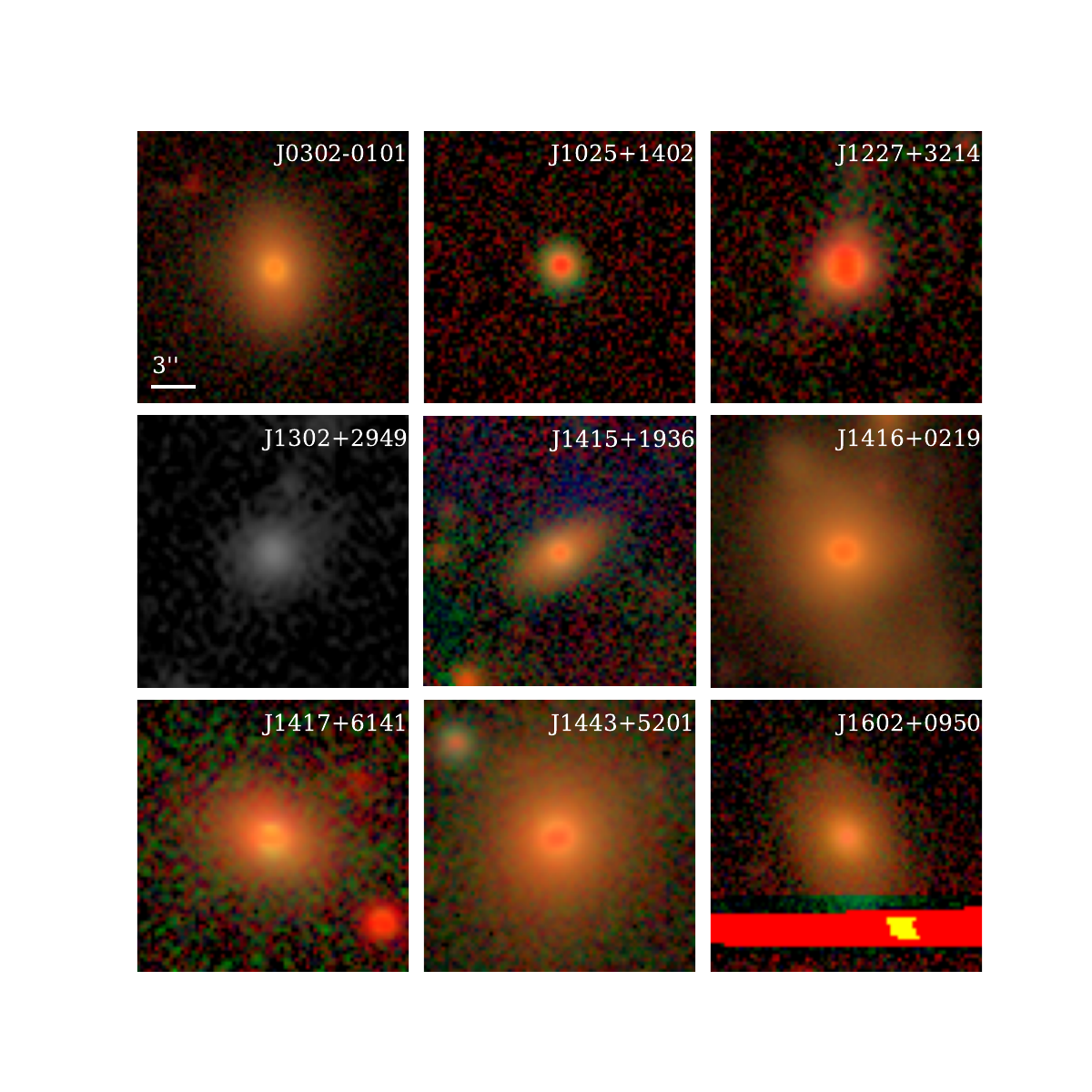}}
	\caption{Legacy Survey DR10 RGB cutouts of the nine local LRD analogues. The cutouts are all 20x20 arcsec.}
	\label{app:photometry}
\end{figure}

Figure \ref{fig:all_spectra_pop_fnu} shows the nine selected objects spectra in $f_\nu$, \ref{app:ztf_sample} presents their ZTF $g$-band light-curves and Fig. \ref{app:photometry} shows Legacy Survey photometric cutouts.

\begin{sidewaystable*}
	\subsection{Parameters of the Nine}\label{app:additional_data}
	In Table \ref{tab:analogues_props} the properties used for the selection criteria are presented.
	In Table \ref{tab:analogues_props_app} additional parameters are listed.
	\caption{Properties of the selected local LRD analogues \textcolor{black}{and the two sources discussed in Sec. \ref{sec:missed_objects}} for the criteria worked out in Sec. \ref{sec:sample_construction}. BBS refers to the Balmer break strength and BD to the Balmer decrement.}
	\label{tab:analogues_props}
	\centering
	\begin{tabular}{ccccccccccc}
		\hline\hline   
		SDSS Name & $\beta_{\mathrm{UV}}$ & $\beta_{\mathrm{OPT}}$ & $\beta_{\mathrm{MBB}}$ & $\lambda_{\mathrm{peak}}$ & BBS &  $L_{\mathrm{H}\alpha}$ &  $L_{5100}$ & [OIII]$_{\lambda 5007}$ EW & $L_{5100}/L_{\mathrm{MBB}}$ &  BD  \\
        & & & & [$\mu$m] & & $[10^{42}\,\mathrm{erg\,s^{-1}}]$ & a & $[\AA]$ & & \\
		\hline
		J0302-0101 & $-1.09 \pm 0.19$ & $0.694 \pm 0.021$ & $-1.91 \pm 0.11$ & $0.66 _{-0.03}^{+0.04}$ & $1.62 _{-0.03}^{+0.03}$ & $2.828 \pm 0.022$ & $53.7 _{-0.8}^{+0.7}$ & $94.9 \pm 0.6$ & $0.23 _{-0.04}^{+0.05}$ & $8.23 \pm 0.28$ \\
		J1025+1402 & $-2.0 \pm 0.6$ & $1.63 \pm 0.04$ & $-1.3 \pm 0.3$ & $0.82 _{-0.09}^{+0.11}$ & $1.50 _{-0.10}^{+0.12}$ & $0.843 \pm 0.005$ & $5.91 _{-0.22}^{+0.20}$ & $261.2 \pm 2.0$ & $0.18 _{-0.06}^{+0.16}$ & $10.9 \pm 0.4$  \\
		J1227+3214 & $-0.73 \pm 0.18$ & $1.985 \pm 0.018$ & $-2.29 \pm 0.11$ & $1.00 _{-0.05}^{+0.06}$ & $1.53 _{-0.03}^{+0.03}$ & $5.93 \pm 0.06$ & $58.4 _{-0.7}^{+0.7}$ & $382.5 \pm 1.6$ & $0.065 _{-0.012}^{+0.019}$ & $10.29 \pm 0.21$ \\
		J1302+2949 & $-1.8 \pm 0.3$ & $0.328 \pm 0.022$ & $-0.34 \pm 0.19$ & $0.56 _{-0.03}^{+0.04}$ & $1.340 _{-0.026}^{+0.029}$ & $2.26 \pm 0.03$ & $47.0 _{-0.8}^{+0.8}$ & $88.2 \pm 2.9$ & $0.53 _{-0.11}^{+0.20}$ & $6.74 \pm 0.27$  \\
		J1415+1936 & $-16 \pm 3$ & $0.123 \pm 0.022$ & $-0.89 \pm 0.19$ & $0.548 _{-0.04}^{+0.04}$ & $1.138_{-0.018}^{+0.018}$ & $1.80 \pm 0.04$ & $34.5 _{-0.7}^{+0.7}$ & $201.8\pm 1.3$ & $0.43 _{-0.09}^{+0.13}$ & $3.76 \pm 0.17$ \\
		J1416+0219 & $-1.2 \pm 0.4$ & $0.614 \pm 0.015$ & $-2.05 \pm 0.08$ & $0.647 _{-0.028}^{+0.028}$ & $1.380 _{-0.016}^{+0.018}$ & $3.936 \pm 0.017$ & $80.4 _{-1.0}^{+1.0}$ & $38.12 \pm 0.28$ & $0.213 _{-0.027}^{+0.04}$ & $5.76 \pm 0.14$  \\
		J1417+6141 & $-1.61 \pm 0.15$ & $0.698 \pm 0.024$ & $-1.49 \pm 0.16$ & $0.64 _{-0.04}^{+0.05}$ & $1.359 _{-0.017}^{+0.017}$ & $3.665\pm 0.015$ & $58.9 _{-0.4}^{+0.4}$ & $28.1\pm 1.1$ & $0.29 _{-0.06}^{+0.10}$ & $5.99\pm 0.16$  \\
		J1443+5201 & $-1.47 \pm 0.19$ & $0.494 \pm 0.018$ & $-1.77 \pm 0.10$ & $0.63_{-0.03}^{+0.03}$ & $1.387 _{-0.018}^{+0.019}$ & $3.287 \pm 0.017$ & $71.0 _{-1.2}^{+1.1}$ & $81.9 \pm 0.5$ & $0.26 _{-0.04}^{+0.05}$ & $4.64 \pm 0.10$  \\
		J1602+0950 & $-1.03 \pm 0.09$ & $0.436 \pm 0.019$ & $-1.34 \pm 0.18$ & $0.61 _{-0.04}^{+0.05}$ & $1.194 _{-0.020}^{+0.020}$ & $4.13 \pm 0.12$ & $91.7 _{-0.9}^{+1.0}$ & $177.0 \pm 1.1$ & $0.35 _{-0.08}^{+0.12}$ & $5.71 \pm 0.23$  \\
		\hline
        \textcolor{black}{J1022+0841} & $-0.84 \pm 0.21$ & $1.78 \pm 0.10$ & $-0.83 \pm 1.27$ & $0.77 _{-0.22}^{+0.5}$ & $1.54 _{-0.29}^{+0.3}$ & $4.22 \pm 0.10$ & $11.2 _{-1.0}^{+1.0}$ & $41.3 \pm 1.5$ & $0.30 _{-0.21}^{+90}$ & $9.7 \pm 0.3$ \\
        \textcolor{black}{J1047+0739} & $-1.0 \pm 0.4$ &  $0.38 \pm 0.05$ & $-4.2 \pm 0.5$ & - & $1.28 _{-0.04}^{+0.05}$ & $4.160 \pm 0.029$ & $19.2 _{-0.7}^{+0.8}$ & $84.0 \pm 0.5$ & - & $5.89 \pm 0.11$  \\
        \hline
	\end{tabular}
	
	\caption{Additional properties of the selected local LRD analogues \textcolor{black}{and the two missed sources discussed in Sec. \ref{sec:missed_objects}}. FWHM$_{\mathrm{H\alpha}}$ is from the \citetalias{Liu19} catalogue, $r_{\mathrm{petro}}$ from SDSS and $\beta_{\mathrm{NIR}}$ we measure from photometry.}
	\label{tab:analogues_props_app}
	\centering
	\begin{tabular}{cccc}
		\hline\hline   
		SDSS Name & FWHM$_{\mathrm{H\alpha}}$ $[\mathrm{km\,s^{-1}}]$ & $r_{\mathrm{petro}}$ [$\mathrm{arcsec}$] & $\beta_{\mathrm{NIR}}$\\
		\hline
		J0302-0101 & $5890\pm 60$ & $3.40 \pm 0.13$ & $0.39 \pm 0.09$ \\
		J1025+1402 & $1342.5 \pm 2.9$ & $1.51 \pm 0.05$ & $0.3 \pm 0.3$ \\
		J1227+3214 & $1613 \pm 27$ & $1.790 \pm 0.015$ & $1.44 \pm 0.04$ \\
		J1302+2949 & $2600 \pm 150$ & $2.40 \pm 0.09$ & $0.38 \pm 0.13$\\
		J1415+1936 & $5100 \pm 100$ & $1.68 \pm 0.04$ & $-0.03 \pm 0.14$ \\
		J1416+0219 & $4050 \pm 50$ & $6.30 \pm 1.19$ & $1.21 \pm 0.05$ \\
		J1417+6141 & $22210 \pm 60$ & $2.98 \pm 0.06$ & $1.00 \pm 0.03$ \\
		J1443+5201 & $5110 \pm 140$ & $5.8 \pm 0.4$ & $0.87 \pm 0.07$ \\
		J1602+0950 & $3930 \pm 250$ & $2.89 \pm 0.17$ & $0.32 \pm 0.09$\\
		\hline
        \textcolor{black}{J1022+0841} & $1260 \pm 27$ & $1.29 \pm 0.07$ & $1.8 \pm 0.3$  \\
        \textcolor{black}{J1047+0739} & $1325 \pm 5$ & $1.51 \pm 0.03$ & $3.23 \pm 0.11$ \\
        \hline
	\end{tabular}
\end{sidewaystable*}


\end{appendix}
	
\end{document}